\documentclass[12pt,a4paper]{article}
\RequirePackage[OT1]{fontenc}
\RequirePackage{amsthm}
\RequirePackage[cmex10]{amsmath}
\RequirePackage{natbib}
\RequirePackage[colorlinks,citecolor=blue,urlcolor=blue, linkcolor=blue, bookmarksopen=true]{hyperref}

\usepackage[margin=1in]{geometry}
\usepackage[latin1]{inputenc}
\usepackage{amsmath}
\usepackage{amsfonts}
\usepackage{amssymb}
\usepackage{comment}
\usepackage{bbm}
\usepackage{graphicx}
\usepackage[usenames,dvipsnames]{color}
\usepackage{multirow}
\usepackage[labelfont=bf]{caption}
\usepackage{placeins}
\usepackage{array}
\usepackage{enumitem}
\usepackage{rotating}
\usepackage{adjustbox}
\usepackage{mathrsfs}
\usepackage{gensymb}
\usepackage{float}
\usepackage{subcaption}
\usepackage{wasysym}
\usepackage[linesnumbered,ruled,lined,boxed,commentsnumbered]{algorithm2e}
\usepackage[linewidth=1pt]{mdframed}
\usepackage{stmaryrd}
\usepackage{thmtools}
\usepackage{thm-restate}
\usepackage{cleveref}

\theoremstyle{definition}

\theoremstyle{plain}

\newtheoremstyle{mytheoremstyle}
    {\topsep}                    
    {\topsep}                    
    {\itshape}                   
    {}                           
    {\scshape}                   
    {:}                         
    {.5em}                       
    {} 
\theoremstyle{mytheoremstyle}

\renewcommand{\restriction}{\mathord{\upharpoonright}}

\SetFuncArgSty{}
\SetProgSty{}
\SetArgSty{}

\title{\textsc{Dividing a cake for the irrationally entitled}\protect}

\author{\textsc{Florian Brandl\thanks{%
Department of Economics, University of Bonn, Bonn, Germany. Email: florian.brandl@uni-bonn.de} and Andrew Mackenzie\thanks{%
Department of Economics, Rutgers University, New Brunswick, NJ, United States of America. Email: andrew.k.mackenzie@rutgers.edu} \hspace{0.01mm}} \thanks{Florian Brandl acknowledges support by the DFG under the Excellence Strategy EXC-2047. We thank \'{A}gnes Cseh, Fedor Sandomirskiy, Erel Segal-Halevi, Tomas Sj\"{o}str\"{o}m, and Warut Suksompong.}}
\date{This draft: \today}

\begin{document}

\maketitle

\begin{abstract}
A perfectly divisible cake is to be divided among a group of agents. Each agent is entitled to a share between zero and one, and these entitlements are compatible in that they sum to one. The mediator does not know the preferences of the agents, but can query the agents to make cuts and appraise slices in order to learn. For each entitlement profile, we provide a lower bound on the number of queries that the mediator requires to construct an allocation that respects the entitlements regardless of preferences. If one of the entitlements is irrational, then the mediator requires an arbitrarily large number of queries.\\


\end{abstract}

\hypertarget{Section1}{}
\section{Introduction}

\hypertarget{Section1.1}{}
\subsection{Overview}

Consider the problem of privatizing some jointly-owned property in accordance with well-defined ownership shares, provided that (i)~dividing the property among the owners is more desirable than selling the property and splitting the proceeds, (ii)~the owners are unable or unwilling to compensate one another using money, and (iii)~whether or not a given allocation respects a given agent's share is a subjective matter that depends on that agent's preferences. For example, a plot of land is to be divided in accordance with a will among heirs who love the old family home, or a divorcing couple's cherished assets are to be divided in accordance with a prenuptial agreement. When the ownership shares are equal, this is the classic problem of fair division \citep{Steinhaus1948}. In this article, we investigate the communication costs associated with solving this problem through mediation when the agents' preferences are private information, building on the contributions of \cite{Lohr2012} and \cite{Cseh-Fleiner2020}. In order to distinguish these communication costs from incentive costs, we proceed under the first-best assumption that the agents will be honest regardless of their incentives, and even so we find that these costs may be unbounded: there may be no deadline by which a solution can be guaranteed.

The importance of considering communication costs when comparing institutions dates back at least to \cite{Hayek1945}, who argued that a crucial benefit of decentralized markets relative to central planning is their lower communication costs. For emphasis, these costs are distinct from other considerations related to private information, such as incentive compatibility, credibility, transparency, and privacy.\footnote{Incentive compatibility is the central consideration in the large literature on mechanism design; see for example \cite{Gibbard1973}. For credibility, see \cite{Akbarpour-Li2020}. For transparency, see for example \cite{Hakimov-Raghavan2023}. For privacy in economics distinct from communication costs, see for example \cite{Gradwohl-Smorodinsky2017}, \cite{Gradwohl2018}, \cite{Milgrom-Segal2020}, \cite{Eliaz-Eilat-Mu2021}, \cite{Mackenzie-Zhou2022}, and \cite{Haupt-Hitzig2024}.} Roughly, the concern we investigate is not that strategically transmitted information may be inaccurate, or that the agents may know too little relative to the mediator throughout the process, or that sensitive information may be unnecessarily exposed; it is instead that mediation may be impractical simply because it may take too long.

There are several ways to measure communication costs. An early definition is the dimensionality of the message space \citep{Hurwicz1977}. In the context of mechanism design, previous work has used {\it communication complexity}, or the worst-case volume of bits (\citealp{Yao1979}; \citealp{Kushilevitz-Nisan1997}), to formalize that incentivizing honesty is costly and that costs can be mitigated by using multi-stage protocols (\citealp{Segal2007}; \citealp{Fadel-Segal2009}; \citealp{Segal2010}). In the context of fair division, costs are measured instead using {\it query complexity}, or the worst-case number of questions asked by the mediator. We show that mediation may be prohibitively expensive even without incentivizing honesty while using a multi-stage protocol, and we do so using query complexity in a model where each query has a one-dimensional set of possible responses. At a high level, the novelty of our contribution can be illustrated using the following two problems.

\vspace{\baselineskip} \noindent \textsc{Problem 1:} A cake must be divided fairly across one million people. It is not enough for the mediator to ensure that each person measures the value of his serving to be at least one-millionth of the cake; the mediator must furthermore ensure that each person deems his serving to be at least as good as any other.

\vspace{\baselineskip} \noindent \textsc{Problem 2:} A decedent leaves a plot of land to be divided across two descendants, a child and a grandchild, in accordance with an unusual will: the child's share of the estate must be equal to the proportion of the grandchild's share to the child's share. The mediator need only ensure that each descendant measures the value of his parcel to be at least his share of the whole plot.

\vspace{\baselineskip} Which problem is harder, as measured by the deadline the mediator can guarantee to the agents, when the only activity that costs any time is communication by means of classic Robertson-Webb queries (\citealp{Robertson-Webb1998}; \citealp{Woeginger-Sgall2007})? Perhaps surprisingly, the two-agent problem is harder. Indeed, for Problem~1, the mediator can guarantee that a solution is found in finite time \citep{Brams-Taylor1995}, and moreover can guarantee a deadline by which it is found \citep{Aziz-Mackenzie2016}. By contrast, for Problem~2, while the mediator can still guarantee that a solution is found in finite time (\citealp{Barbanel1995}; \citealp{Shishido-Zeng1999}; \citealp{Cseh-Fleiner2020}; \citealp{Janko-Joo2022}), we show that the mediator can never guarantee a deadline if there is an agent whose entitlement is an irrational number (\hyperlink{Corollary1}{Corollary~1}), and this is the case for Problem~2.\footnote{Indeed, the child's share $x \in [0,1]$ must satisfy $ 1-x = x^2$, or $x = \frac{-1+\sqrt{5}}{2} \approx 61.8\%$; this is simply the inverse of the golden ratio.} Notably, this establishes the existence of a classic cake division problem that is finite but unbounded.

\hyperlink{Corollary1}{Corollary~1}, in turn, follows from our main result, which involves associating each entitlement profile with an index familiar to the literature: the least common denominator of the entitlements, defined as infinity when there is an irrational entitlement. We refer to this as the {\it clonage} to suggest the minimum number of clones required to form an economy with equal entitlements, provided we must replace each agent with some number of clones who equally split his entitlement.\footnote{There is no word in English for {\it amount of clones}. We propose {\it clonage}, like coverage for an amount covered, leakage for an amount leaked, or mileage for an amount of miles, but unlike coinage for the creation of a new word.}

It is an old observation that we can solve a problem with finite clonage by solving the associated clone problem where each agent's clones inherit his preferences, and that this in turn can be solved by a guaranteed deadline. Indeed, the idea can be illustrated simply using the Dubins-Spanier variant of the Banach-Knaster protocol (\citealp{Steinhaus1948}; \citealp{Dubins-Spanier1961}): given a knife that glides over the cake from left to right, (i)~ask each of the $n$ clones to make a cut such that he measures the value to the left of his cut to be~$\frac{1}{n}$, (ii)~give a clone with a left-most cut everything to the left of his cut, and (iii)~iterate. We solve the original problem by giving each agent all slices assigned to his clones, and if the original entitlement profile has clonage $n$, then we require $\sum_{i=1}^n i = \frac{n^2+n}{2}$ queries.

It is far less obvious, but true, that it is unnecessarily costly for the mediator to solve the associated clone problem. Indeed, costs can be reduced by using the more elaborate Even-Paz protocol \citep{Even-Paz1984} in place of the simple Dubins-Spanier protocol, and in a suitable sense this is the optimal protocol for solving the clone problem \citep{Edmonds-Pruhs2011}. That said, \cite{Cseh-Fleiner2020} show that costs can be reduced {\it further} by directly solving the original problem using a recursive version of the two-agent Cut Near-Halves protocol \citep{Robertson-Webb1998}, and that costs can be reduced {\it further still} by iteratively splitting the original problem into two sub-problems using the Cseh-Fleiner protocol \citep{Cseh-Fleiner2020}.

Even so, clonage remains relevant to the analysis. In particular, if there are $n$ agents and the clonage level is $\mathsf{c}$, then the Cseh-Fleiner protocol requires $2 (n-1) \lceil \log_2 \mathsf{c} \rceil$ queries \citep{Cseh-Fleiner2020}. Curiously, then, even though the Cseh-Fleiner protocol does not solve the associated clone problem, it grows more costly with clonage. Why?

Our main result provides a general answer: even though it is unnecessarily costly to solve the associated clone problem, {\it clonage causes communication costs}. More precisely, for each clonage level $\mathsf{c}$, and for each problem with~$n$ agents and a clonage above $\mathsf{c}$, {\it no} protocol is guaranteed to finish within $\lfloor \log_2 \log_2 2 \mathsf{c}^{\frac{1}{n-1}} \rfloor$ queries (\hyperlink{Theorem1}{Theorem~1}). Our lower bound and the Cseh-Fleiner upper bound together establish that a sufficient increase in clonage alone can cause the mediator's problem to become harder, though the gap between these bounds leaves open the question of whether clonage is responsible for {\it all} communication costs.

The explanation relies on an alternative interpretation of clonage: instead of viewing it as an index of how many {\it agent} clones we require, we view it as an index of how many {\it cake} clones we require. More precisely, the clonage $\mathsf{c}$ is the minimum number of miniature cake replicas required to solve the problem in zero queries, provided that each agent values each serving of a replica to be worth $\frac{1}{\mathsf{c}}$ of the associated serving in the original cake. This is an index of the difficulty the mediator faces at his initial information set, and we establish our main result by generalizing this difficulty index to other information sets.

\hypertarget{Section1.2}{}
\subsection{Related literature}

To facilitate this discussion, we first set the stage with some key definitions in the context of a classic concrete example.

\vspace{\baselineskip} \noindent \textsc{Example 1:} The mediator must allocate the {\it interval cake} $C = (0, 1]$ using the {\it standard knife} $(\kappa_t)_{t \in [0, 1]}$ defined as follows: for each $t \in [0, 1]$, we have $\kappa_t = (0, t]$. A {\it serving} is any finite union of intervals of the form $(a, b]$, and a {\it kitchen measure} is a probability measure that assigns a positive value to each nonempty serving. The set of {\it agents} is $N = \{1, 2, ..., n\}$, the {\it entitlement profile} is $e \in [0,1]^N$ with $\sum_{i \in N} e_i = 1$, and an {\it allocation} is an assignment of servings to agents $X = (X_i)_{i \in N}$ such that the servings are pairwise disjoint and collectively exhaustive. Initially, the mediator knows that the agents' preferences are represented by a kitchen measure profile $\mu = (\mu_i)_{i \in N}$, but does not know which. In the context of a kitchen measure profile $\mu$, an allocation $X$ is (i)~{\it proportional} if for each $i \in N$ we have $\mu_i(X_i) \geq \frac{1}{n}$, (ii)~{\it envy-free} if for each pair $i, j \in N$ we have $\mu_i(X_i) \geq \mu_i(X_j)$, and (iii)~{\it $e$-proportional} if for each $i \in N$ we have $\mu_i(X_i) \geq e_i$.

\hypertarget{Table1}{}
\begin{table}
\footnotesize
\centering
\begin{tabular}{c | c | c}
axiom & lower bound & upper bound
\\ \hline \rule{0pt}{4ex} no-envy & $\Omega(n^2)$, \citealp{Procaccia2009} & $\mathcal{O}\Big(n^{n^{n^{n^{n^n}}}}\Big)$, \citealp{Aziz-Mackenzie2016}\footnotemark
\\ proportionality & $\Omega(n \log_2 n)$, \citealp{Edmonds-Pruhs2011} & $\mathcal{O}(n \log_2 n)$, \citealp{Even-Paz1984}
\\ $e$-proportionality & $\Omega(\log_2 \log_2 \mathsf{C}(e))$, \hyperlink{Corollary2}{Corollary~2} & $\mathcal{O}(\log_2 \mathsf{C}(e))$, \citealp{Cseh-Fleiner2020}
\end{tabular}
\caption{\footnotesize {\it Query bounds for cake division.} We hold the cake and knives fixed. The number of agents is~$n$, the clonage of entitlement profile $e$ is $\mathsf{C}(e)$, and for each problem the $\mathsf{cost}$ we bound is the worst-case number of queries across measure profiles. In the first two rows, $n$ alone determines a unique problem. In the third row, an arbitrary number of agents $n \geq 2$ is held constant, and the clonage determines a class of problems indexed by entitlement profiles; the lower bound is for the best-case $\mathsf{cost}$ across problems given the clonage and the upper bound is for the worst-case $\mathsf{cost}$ across problems given the clonage. See \protect\hyperlink{Section4.2}{Section~4.2} for the formal definitions.}
\end{table}
\footnotetext{Some recent working papers report improvements to this bound (\citealp{Sokolov2023}; \citealp{Ye-Bai2026}).}

\vspace{\baselineskip} For each entitlement profile $e$, we establish a lower bound on the worst-case number of {\it queries} that the mediator must ask in an adaptive multi-stage {\it protocol} that determines an $e$-proportional allocation regardless of the kitchen measure profile. Our lower bound is established with particularly powerful queries, and thus holds for the classic Robertson-Webb queries used throughout this discussion; see \cite{Cseh-Fleiner2020} for a detailed comparison of other query models.

We begin by discussing query bounds for envy-free allocation, proportional allocation, and $e$-proportional allocation, in sequence; we then conclude with some remarks about contributions we make to the model. Recall that in asymptotic order notation, $\mathcal{O}(f(x))$ is an upper bound and $\Omega(f(x))$ is a lower bound; see \hyperlink{Section4.2}{Section~4.2} for the familiar definitions.

\vspace{\baselineskip} \noindent \textsc{No envy.} When there are two agents, a protocol used by Prometheus and Zeus in Hesiod's {\it Theogony}, which dates from the eighth or seventh century BC (see \citealp{Lowry1987}; \citealp{Brams-Taylor-Zwicker1995}), constructs an allocation that is not only proportional but moreover envy-free: one agent divides the cake into two parts he deems equal, then the other agent chooses. In our model, because we work with particularly powerful queries, this is a one-query protocol.

Calls for a three-agent envy-free protocol date to at least \cite{Gamow-Stern1958}, and shortly thereafter, such a protocol was independently discovered by John Selfridge and John Conway; see \cite{Brams-Taylor1995}. The extension of the Selfridge-Conway protocol to even four agents remained a major open question for decades, until the arrival of the Brams-Taylor protocol for any number of agents \citep{Brams-Taylor1995}.

For each kitchen measure profile, the Brams-Taylor protocol uses a finite number of queries, and yet for each deadline the mediator might hope to guarantee, there is a kitchen measure profile that takes too long. To quote \cite{Procaccia2020}: ``Consequently, as soon as Brams and Taylor solved the envy-free cake-cutting problem, they immediately launched a new problem to the top of fair division's most wanted list: the existence of a bounded envy-free cake-cutting protocol." In the meantime, it was established that (i)~any envy-free protocol requires at least $\Omega(n^2)$ queries \citep{Procaccia2009}, which together with the Even-Paz protocol for proportional allocation \citep{Even-Paz1984} established that envy-free allocation is more difficult than proportional allocation, and (ii)~even though the interval cake always has an envy-free allocation for which each agent consumes an interval (\citealp{Stromquist1980}; \citealp{Woodall1980}), there is no finite protocol to find such an allocation \citep{Stromquist2008}.

Finally, after enjoying decades as the leading candidate for a classic cake division problem that is finite but unbounded, the problem of envy-free allocation was ultimately classified as bounded with the arrival of the Aziz-Mackenzie protocol \citep{Aziz-Mackenzie2016}. We continue the conversation by classifying a {\it different} classic cake division problem as finite but unbounded, establishing that just as envy-free allocation is more difficult than proportional allocation, so too is $e$-proportional allocation more difficult than envy-free allocation, with envy-free allocation of the interval cake using only intervals harder still.

\vspace{\baselineskip} \noindent \textsc{Rational entitlements.} In the seminal contribution to fair division, \cite{Steinhaus1948} introduced his own proportional protocol for three agents as well as the Banach-Knaster protocol for any number of agents. As discussed earlier, the latter has a particularly simple variant involving a moving knife \citep{Dubins-Spanier1961}, yielding the original upper bound for proportional allocation of $\mathcal{O}(n^2)$. This was then improved to $\mathcal{O}(n \log_2 n)$ by the Even-Paz protocol \citep{Even-Paz1984}, which after some partial lower bound results (\citealp{Magdon-Ismail-Busch-Krishnamoorthy2003}; \citealp{Woeginger-Sgall2007}) was shown to be optimal \citep{Edmonds-Pruhs2011}.

The seminal contribution also remarked that the Banach-Knaster protocol can be adapted to rational shares and mused about protocol performance. To quote \cite{Steinhaus1948}: ``The procedure described above applies also, under slight modifications, to the case of different (but rational) ideal shares. Interesting mathematical problems arise if we are to determine the minimal numbers of `cuts' necessary for fair division." Presumably, the slight modifications refer to solving the associated clone problem discussed earlier.

Decades later, two alternatives to cloning were proposed for two-agent problems: the Ramsey Partition protocol \citep{McAvaney-Robertson-Webb1992} and the Cut Near-Halves protocol \citep{Robertson-Webb1998}. The latter requires fewer queries than the former \citep{Robertson-Webb1998}, and a recursive generalization of the latter requires fewer queries than solving the associated clone problem with {\it its} optimal protocol \citep{Cseh-Fleiner2020}. Decades later still, the Cseh-Fleiner protocol was proposed as a further improvement, and it provides the current upper bound on the number of queries required for $e$-proportional allocation \citep{Cseh-Fleiner2020}. As discussed earlier, even though the Cseh-Fleiner protocol does not solve the associated clone problem, its upper bound grows with clonage.

To our knowledge, the first lower bound was provided by \cite{Lohr2012}. Lohr's lower bound concerns dividing the interval cake for two agents when the mediator is constrained to use a protocol that is {\it simple}, in the sense that after each cut, one of the two pieces must be immediately assigned on the basis of whether or not the agent who did not cut values the cut piece more than its target value. There is reason to suspect the simplicity constraint imposes significant costs on the mediator: when one of the entitlements is irrational, there are finite protocols (\citealp{Barbanel1995}; \citealp{Shishido-Zeng1999}; \citealp{Cseh-Fleiner2020}; \citealp{Janko-Joo2022}), but it is easy to see that no simple protocol is finite. Even so, our key proposition (\hyperlink{Proposition1}{Proposition~1}) shows that Lohr's lower bound persists when the mediator is allowed to use any protocol. Lohr {\it also} shows that this lower bound is in a suitable sense tight; see \hyperlink{Section4}{Section~4} for details.

Finally, \cite{Cseh-Fleiner2020} provide a lower bound that allows for a general class of protocols, and its distinction from our lower bound is subtle but significant. In particular, Cseh-Fleiner provide a lower bound for each problem as a function of what we call its {\it fineness}: the smallest $n \in \mathbb{N}$ such that $\frac{1}{n}$ is at most the smallest positive entitlement.\footnote{Intriguingly, tiny entitlements have also proven troublesome for a mechanism design problem related to privatizing jointly-owned property in accordance with well-defined ownership shares. In particular, if two agents with private preference information are to allocate one indivisible object and monetary transfers are feasible, then the entitlement profile has the property that it is possible to implement efficient and voluntary allocations if and only if no agent has a tiny entitlement \citep{Cramton-Gibbons-Klemperer1987}. (More generally, for more than two agents, no agent should have a huge entitlement.) Here, even zero counts as a problematic tiny entitlement because an agent who owns nothing can still pay; thus the classic \cite{Myerson-Satterthwaite1983} impossibility theorem is a special case.} This can be rephrased using clonage in the following sense: if for each clonage level we cherry-pick a problem that is as fine as possible, then the number of required queries is $\Omega(\log_2 \mathsf{C}(e))$, where $\mathsf{C}(e)$ denotes the clonage of $e$. By contrast, we prove that if for each clonage level we cherry-pick a problem that is as {\it easy} as possible, then the number of required queries is $\Omega(\log_2 \log_2 \mathsf{C}(e))$. These lower bounds are not logically related, and at first glance they may appear rather similar. That said, the strong conclusions that we draw in our overview require our lower bound. Indeed, for Problem 2, where the mediator must divide the decedent's estate across his child and grandchild, the fineness level is three but the clonage level is infinity. See \hyperlink{Section4}{Section~4} for further discussion.

\vspace{\baselineskip} \noindent \textsc{General entitlements.} The general problem of constructing an $e$-proportional allocation was first considered by \cite{Barbanel1995}, and to date there are five finite protocols that solve this problem: the Barbanel protocol \citep{Barbanel1995}, the Shishido-Zeng protocol \citep{Shishido-Zeng1999}, the {\it second} Cseh-Fleiner protocol \citep{Cseh-Fleiner2020}, and the two Jank\'{o}-Jo\'{o} protocols \citep{Janko-Joo2022}. All five protocols are unbounded, and we provide a simple explanation: {\it no} protocol that solves this problem is bounded.

\vspace{\baselineskip} \noindent \textsc{The model.} First, we provide a measure-theoretic approach to cake division protocols that to our knowledge is novel. Though measure-theoretic cakes are not new \citep{Dubins-Spanier1961}, previous contributions to cake division protocols generally assert that the cake lives in Euclidean space alongside the Lebesgue measure, then use the Lebesgue measure to construct knives and articulate the associated preference requirements (\citealp{Woeginger-Sgall2007}; \citealp{Segal-Halevi-Nitzan-Hassidim-Aumann2020}; \citealp{Cseh-Fleiner2020}). By contrast, we directly specify the knives as the primitives from which the servings are constructed, then use the knives to articulate our preference requirements. This additional generality is largely an aesthetic choice, guided by the subjective principle that unnecessary details are often distracting.

Second, we provide a preference foundation for kitchen measures. While there are existing foundations for the measures required for cake division \citep{Barbanel-Taylor1995}, we are able to use the additional structure that our algebra of servings inherits from our primitive knives to provide foundations that are closer to classic decision theory. In particular, ranking cake servings on the basis of preferences is mathematically equivalent to ranking events on the basis of beliefs, and it is already known that Savage's classic foundation for representing beliefs with measures \citep{Savage1954} extends from power sets to algebras (\citealp{Wakker1981}; \citealp{Marinacci1993}). In order to obtain kitchen measure representation, we further impose that (i)~a serving is null if and only if it is what we call a {\it sliver}, and (ii)~preferences are continuous with respect to knives. The contribution is that with our additional assumptions, we are able to drop Savage's technical {\it tightness} axiom, replacing it with two conditions that are somewhat easier to interpret (\hyperlink{Theorem2}{Theorem~2}).

Finally, deviating from earlier literature, we include a formal definition of protocols; they are strategies for the mediator in a particular one-player extensive-form game against nature. This formality allows us to translate observations into the language of game theory, including notably our observation of an interesting phenomenon in a variant two-player zero-sum game between the mediator and the {\it adversary}: when we restrict attention to pure strategies, the game has a value but no Nash equilibrium.

\hypertarget{Section2}{}
\section{Problems}

\hypertarget{Section2.1}{}
\subsection{Kitchens}

In our problem, a mediator must cut a cake into servings while constrained by the available technology for making cuts.

\vspace{\baselineskip} \noindent \textsc{Definition:} A {\it kitchen} is a pair $(C, \mathbb{K})$ that satisfies the following.
\begin{itemize}
\item $C$ is a nonempty set referred to as the {\it cake}.

\item A {\it knife} is a strictly monotonic function $\kappa: [0, 1] \to 2^C$, where for each $t \in [0, 1]$ we write $\kappa_t$ instead of $\kappa(t)$. Equivalently, $(\kappa_t)_{t \in [0,1]} \in (2^C)^{[0,1]}$ is a knife if for each pair $t, t' \in [0, 1]$ such that $t' > t$, we have $\kappa_t \subsetneq \kappa_{t'}$.

\item $\mathbb{K}$ is a nonempty set of knives referred to as the {\it knife block}. This represents the available technology for cutting the cake.

\item A {\it simple slice} is a member of $\cup_{\kappa \in \mathbb{K}} \cup_{t \in [0,1]} \{ \kappa_t, C \backslash \kappa_t \}$, and a {\it slice} is any finite intersection of simple slices. We let $\mathcal{S}^\cap \subseteq 2^C$ denote the collection of slices. By convention, we include $C \in \mathcal{S}^\cap$ as the slice associated with the empty collection of simple slices.

\item A {\it serving} is a finite union of slices. We let $\mathcal{S} \subseteq 2^C$ denote the collection of servings. By convention, we include $\emptyset \in \mathcal{S}$ as the serving associated with the empty collection of slices. Observe that each slice is a serving: $\mathcal{S}^\cap \subseteq \mathcal{S}$.
\end{itemize}
Observe that $\mathcal{S}$, together with its set-theoretic structure, is a Boolean algebra.\footnote{More precisely, $\mathcal{S}$ is the subalgebra of $2^C$ generated by the simple slices. It is clear that the latter contains the former, and to see that the former contains the latter it suffices to write each member of the latter in disjunctive normal form.}

\vspace{\baselineskip} We remark that the model has several alternative interpretations---for example, a slice might represent a parcel of land, a time slot on a schedule, or an uncertain event---but for brevity we stick to the cake interpretation.

Intuitively, if knife $\kappa$ glides over the cake from time zero to time one, then it can make a cut by stopping at any time $t \in [0, 1]$, in which case the cake is split into $\kappa_t$ and $C \backslash \kappa_t$. We will eventually require that agents declare $\kappa_0$ to be equivalent to no cake while declaring $\kappa_1$ to be equivalent to the entire cake, so it may already be useful to imagine that a knife glides across (essentially) the entire cake. For our problem, the construction of a particular serving may involve multiple knives, but we assume that the technology for constructing servings is constrained to the fixed inventory of knives given by the knife block, and as a result some subsets of the cake may not be feasible servings.

\vspace{\baselineskip} \noindent \textsc{Example 1 (Revisited):} Let $C = (0, 1]$; we refer to this as the {\it interval cake}. Define the {\it interval knife} $(\kappa_t)_{t \in [0, 1]}$ as follows: for each $t \in [0, 1]$, $\kappa_t \equiv (0, t]$. Finally, define the knife block $\mathbb{K} \equiv \{ (\kappa_t)_{t \in [0, 1]} \}$. We refer to $(C, \mathbb{K})$ as the {\it interval cake kitchen}. This is our model's formalization of the classic interval cake discussed in \hyperlink{Section1.2}{Section~1.2}.

\vspace{\baselineskip} \noindent \textsc{Example 2:} Let $C = (0, 1]^2$; we refer to this as the {\it square cake}. Define the {\it standard vertical knife} (which moves horizontally), $(\kappa_t)_{t \in [0, 1]}$, and the {\it standard horizontal knife} (which moves vertically), $(\kappa'_t)_{t \in [0, 1]}$, as follows: for each $t \in [0, 1]$, $\kappa_t \equiv \{(x, y) \in C | x \leq t \}$ and $\kappa'_t \equiv \{(x, y) \in C | y \leq t \}$. Finally, define the knife block $\mathbb{K} \equiv \{ (\kappa_t)_{t \in [0, 1]}, (\kappa'_t)_{t \in [0, 1]} \}$. We refer to $(C, \mathbb{K})$ as the {\it (standard) square cake kitchen}. In this case, $\{(x, y) \in C | x = y\}$ is a simple example of a subset of the cake that is not a serving.

\vspace{\baselineskip} Before proceeding, we observe that we allow for algebras that are not $\sigma$-algebras. This point is not crucial to our main message, but it is relevant to some detailed remarks about our relationship with the literature, and this point can be illustrated using our first example.

\vspace{\baselineskip} \noindent \textsc{Example 1 (Revisited):} Let $(C, \mathbb{K})$ be the interval cake kitchen. We show that a countable family of servings may have a supremum in $\mathcal{S}$ that is distinct from its supremum in $2^C$, and may also have no supremum in $\mathcal{S}$. Indeed, for each $m \in \mathbb{N}$,\footnote{In this paper, we write $\mathbb{N} = \{1, 2, ...\}$ and $\mathbb{N}_0 = \mathbb{N} \cup \{0\}$.} define $S_m \equiv (0, 1 - \frac{1}{2^m}]$ and $S'_m \equiv \cup_{m' \in \{1, 2, ..., m\}} (\textstyle{\frac{1}{3^{m'}}}, \frac{2}{3^{m'}}]$. First, the family $\{S_m\}_{m \in \mathbb{N}}$ has supremum $(0, 1)$ in $2^C$ and supremum $(0, 1]$ in~$\mathcal{S}$; these are distinct and the former does not belong to $\mathcal{S}$. Second, the family $\{S'_m\}_{m \in \mathbb{N}}$ has no supremum in $\mathcal{S}$; thus $\mathcal{S}$ is not a $\sigma$-algebra.

\hypertarget{Section2.2}{}
\subsection{Kitchen measures}

The cake will ultimately be consumed by a group of agents, each agent has preferences over servings, and we make assumptions about these preferences that imply they can be represented by exactly one measure with some additional structure. In this case, we say that the measure assigns values to servings, and we use these values to articulate whether or not a given allocation is satisfactory.

In our problem, the mediator does not know the agents' preferences. To make the mediator's problem easier---and thus to make our negative result stronger---we suppose that the mediator knows that for each agent, a slice is worth nothing if and only if it is what we call a {\it sliver}: a slice that some knife can cut no further, or a serving contained in such a slice, or a finite union of such servings.

\vspace{\baselineskip} \noindent \textsc{Definition:} Fix a kitchen. We define slivers in three steps: (i) a slice $S \in \mathcal{S}^\cap$ is a {\it slice-sliver} if there is $\kappa \in \mathbb{K}$ such that for each $t \in [0, 1]$, we have $S \cap \kappa_t = \emptyset$ or $S \backslash \kappa_t = \emptyset$, (ii)~a {\it subslice-sliver} is a serving that is a subset of a slice-sliver, and (iii)~a {\it sliver} is a finite union of subslice-slivers. We let $\mathcal{S}^\circ \subseteq \mathcal{S}$ denote the collection of slivers; this is the ideal generated by the slice-slivers. By convention, we include $\emptyset \in \mathcal{S}^\circ$ as the sliver associated with the empty collection of subslice-slivers.

\vspace{\baselineskip} In order to learn about preferences, the mediator may select an agent, a knife, a serving, and a target proportion, then request that the agent use the knife to cut a part from the serving such that the part's value is the target proportion of the serving's value. To make the mediator's problem even easier---and thus to make our negative result even stronger---we suppose that the mediator knows that agents are always able to complete these requests. In particular, we assume that the mediator knows that each agent's preferences are represented by what we call a {\it kitchen measure}. We remark that for another cake division problem, analogous knowledge has been shown to be useful to the mediator \citep{Janko-Joo-Segal-Halevi-Yuen2026}.\footnote{In particular, for the problem of partitioning the interval cake such that each agent receives an interval worth strictly more than his entitlement, when such an allocation is known to exist, the mediator can do better when he knows that all agents are {\it hungry}, or that no agent considers any positive interval to be worthless \citep{Janko-Joo-Segal-Halevi-Yuen2026}.}

\vspace{\baselineskip} \noindent \textsc{Definition:} Fix a kitchen. A {\it probability measure} is a function $\mu_0: \mathcal{S} \to [0, 1]$ such that (i)~$\mu_0(C) = 1$, and (ii)~for each pair $A, B \in \mathcal{S}$ such that $A \cap B = \emptyset$, we have $\mu_0(A \cup B) = \mu_0(A) + \mu_0(B)$. In this case, we say that $\mu_0$ is moreover a {\it kitchen measure} if it satisfies the following conditions.
\begin{itemize}
\item {\it Null slivers.} For each $S \in \mathcal{S}$, we have $\mu_0(S) = 0$ if and only if $S \in \mathcal{S}^\circ$.

\item {\it Knife divisibility.} For each $A \in \mathcal{S}$, each $p \in [0, 1]$, and each $\kappa \in \mathbb{K}$, there is $t \in [0, 1]$ such that $\mu_0(A \cap \kappa_t) = p \cdot \mu_0(A)$.
\end{itemize}
We let $\mathbb{M} \subseteq [0, 1]^\mathcal{S}$ denote the collection of kitchen measures. We say that $(C, \mathbb{K})$ is {\it tidy} if and only if $\mathbb{M} \neq \emptyset$.

\vspace{\baselineskip} To conclude this section, we first illustrate the definition of kitchen measure using the square cake, then discuss tidy kitchens, and finally discuss preference assumptions. We begin with the illustration.

\vspace{\baselineskip} \noindent \textsc{Example 2 (Revisited):} Let $(C, \mathbb{K})$ be the square cake kitchen. The restriction of the Lebesgue measure to $\mathcal{S}$ is a kitchen measure, so the square cake kitchen is tidy. The only sliver is the empty serving, so in order for a probability measure to be a kitchen measure, null slivers requires that each nonempty serving is assigned a positive value.

To see that null slivers does not imply knife divisibility, let $V$ denote the vertical line segment $\{0.5\} \times [0, 1]$. For emphasis, $V$ is not a serving and thus not a sliver. For each $S \in \mathcal{S}$, (i)~define $\mu^*(S)$ to be the one-dimensional Lebesgue measure of $S \cap V$, (ii)~define $\mu^{**}(S)$ to be the two-dimensional Lebesgue measure of $S$, and (iii)~define $\mu(S) \equiv \frac{1}{2}\mu^*(S) + \frac{1}{2}\mu^{**}(S)$. Then~$\mu$ is a probability measure that satisfies null slivers, yet it violates knife divisibility: for the standard vertical knife $\kappa$, there is no $t \in [0, 1]$ such that $\mu(C \cap \kappa_t) = \frac{1}{2} \cdot \mu(C) = \frac{1}{2}$.

To see that knife divisibility does not imply null slivers, define $C^* \equiv (0, \frac{1}{2}] \times (0, \frac{1}{2}]$, and for each $S \in \mathcal{S}$, define $\mu(S)$ to be the two-dimensional Lebesgue measure of $S \cap C^*$ multiplied by four. Then $\mu$ is a probability measure that satisfies knife divisibility, yet it violates null slivers because $(\frac{1}{2}, 1] \times (0, 1]$ is not a sliver but nevertheless is assigned zero value.

\vspace{\baselineskip} As the previous example illustrates, the square kitchen is tidy. As the next example illustrates, not all kitchens are tidy.

\vspace{\baselineskip} \noindent \textsc{Example 3:} Let $C = (0, 1]^\mathbb{R}$, and for each dimension $d \in \mathbb{R}$, let $\kappa^d$ be the knife such that for each $t \in [0, 1]$,
\begin{align*}
\kappa^d_t = \{x \in C | x_d > 1-t \text{ and for each } d' \in \mathbb{R} \backslash \{d\} \text{ we have } x_{d'} \leq t\}.
\end{align*}
For each $d \in \mathbb{R}$, define $A_d \equiv \kappa^d_{\frac{1}{2}}$. Then $\{A_d\}_{d \in \mathbb{R}}$ is a pairwise disjoint continuum of servings that are not slivers, so there cannot be a kitchen measure. Indeed, assume by way of contradiction that $\mu \in \mathbb{M}$, and for each $m \in \mathbb{N}$, define $\mathcal{D}_m \equiv \{ d \in \mathbb{R} | \mu(A_d) \in (\frac{1}{m+1}, \frac{1}{m}]\}$. Then (i)~$\cup_{m \in \mathbb{N}} \mathcal{D}_m = \mathbb{R}$, so $|\cup_{m \in \mathbb{N}} \mathcal{D}_m| = |\mathbb{R}|$, and (ii)~for each $m \in \mathbb{N}$ we have $|\mathcal{D}_m| \leq m$, so $|\cup_{m \in \mathbb{N}} \mathcal{D}_m| \leq |\mathbb{N}|$. But then $|\mathbb{R}| \leq |\mathbb{N}|$, contradicting $|\mathbb{N}| < |\mathbb{R}|$.

\vspace{\baselineskip} In order for a kitchen to be tidy, the quotient algebra---obtained by declaring two servings to be equivalent if their symmetric difference is a sliver, gathering the equivalence classes, and deriving algebraic operations from the original algebra in the natural way---must carry a strictly positive measure and therefore satisfy Kelley's criterion \citep{Kelley1959}. The cake division literature has largely focused on tidy kitchens obtained like the interval cake kitchen or the square cake kitchen: begin with a probability space equipped with a countably additive and atomless reference measure---usually Euclidean space equipped with the Lebesgue measure---and then require each knife's nested family of servings to be strictly monotonic in the reference measure. We remark that the inquiry into when certain algebras are compatible with certain measures traces back to at least 1937, when John von Neumann raised Problem~163 of the Scottish Book \citep{Mauldin1981}.

We conclude this section by discussing preference assumptions. To reiterate, we make assumptions about each agent's preferences that imply they can be represented by exactly one kitchen measure. What, exactly, are these assumptions? The short answer is that we are assuming (i)~the classic qualitative probability axioms (\citealp{Bernstein1917}; \citealp{deFinetti1937}; \citealp{Koopman1940}), (ii)~one of the two technical axioms of \cite{Savage1954}, (iii)~the null servings are the slivers, and (iv)~each knife is continuous in a suitable sense. Notably, these assumptions together imply the second technical axiom of \cite{Savage1954}. These claims require further elaboration, but as they are not crucial to our main result about communication costs, we postpone the details until \hyperlink{Section5.1}{Section~5.1}.

\hypertarget{Section2.3}{}
\subsection{Settings and entitlements}

The cake will ultimately be consumed by a group of agents, and the mediator must partition the cake into servings and then match these servings to the agents so that the resulting allocation is, in a sense to be made precise shortly, satisfactory.

\vspace{\baselineskip} \noindent \textsc{Definition:} A {\it setting} is a tuple $(C, \mathbb{K}, N)$ that satisfies the following.
\begin{itemize}
\item $(C, \mathbb{K})$ is a tidy kitchen.

\item $N$ is a finite and nonempty set of {\it agents}, with $n \equiv |N|$.

\item $\mathcal{X} \subseteq \mathcal{S}^N$ is the set of {\it allocations}: for each $X \in \mathcal{S}^N$, we have $X \in \mathcal{X}$ if and only if (i)~for each distinct pair $i, j \in N$, we have $X_i \cap X_j = \emptyset$, and (ii)~$\cup_{i \in N} X_i = C$.
\end{itemize}
All of our results involve fixing an arbitrary setting.

\vspace{\baselineskip} In our problem, each agent is equipped with both (i)~an entitlement to a share of the cake between zero and one, and (ii)~a kitchen measure representing his preferences. Moreover, the agents' entitlements are compatible in that they sum to one. Whether or not a given allocation is satisfactory depends on both the entitlements and the kitchen measures: each agent should receive a serving that is worth at least his entitlement according to his measure.

\vspace{\baselineskip} \noindent \textsc{Definition:} Fix a setting.
\begin{itemize}
\item $E \equiv \{ e \in [0, 1]^N | \sum e_i = 1\}$ is the set of {\it entitlement profiles}. 

\item $\mathbb{D} \equiv \mathbb{M}^N$ is the {\it unrestricted domain (of all kitchen measure profiles)}.
\end{itemize}
For each $e \in E$ and each $\mu \in \mathbb{D}$, an allocation $X \in \mathcal{X}$ is {\it $(e|\mu)$-proportional} if for each $i \in N$, we have $\mu_i(X_i) \geq e_i$. When omitting $\mu$ creates no confusion, we use the simpler term {\it $e$-proportional}.

\vspace{\baselineskip} In the special case that each agent has entitlement $\frac{1}{n}$, $e$-proportionality coincides with the classic notion of proportionality \citep{Steinhaus1948}. As discussed earlier, the general existence of $e$-proportional allocations has been shown by construction (\citealp{Barbanel1995}; \citealp{Shishido-Zeng1999}; \citealp{Cseh-Fleiner2020}; \citealp{Janko-Joo2022}), and there are particularly simple proofs of existence for two special cases: (i)~each agent's measure is countably additive (\citealp{Liapounoff1940}; \citealp{Dubins-Spanier1961}), and (ii)~each entitlement is a rational number (\citealp{Steinhaus1948}; \citealp{Dubins-Spanier1961}).

We consider a mediator who knows the entitlement profile but not the kitchen measure profile, and who is tasked with constructing an $e$-proportional allocation {\it regardless of the kitchen measure profile}. To do so, the mediator may communicate with the agents in order to learn more about their preferences, and the allocation that the mediator constructs may be contingent on the information that he receives. That said, communication is costly, and the mediator's problem is to keep these costs down.

\hypertarget{Section3}{}
\section{Solutions}

\hypertarget{Section3.1}{}
\subsection{Queries}

In order to learn about the agents' preferences, the mediator selects a {\it query} and receives a {\it response}. Since the seminal contribution of \cite{Robertson-Webb1998}, these terms have been formalized in several different ways (\citealp{Woeginger-Sgall2007}; \citealp{Edmonds-Pruhs2011}; \citealp{Cseh-Fleiner2020}).

Our choice of definition is based on two considerations. First, as an aesthetic choice, we require that whenever a cut is made, the mediator does not lose access to old queries or gain access to new queries: the mediator adaptively selects from the fixed set of queries. Second, because our main result loosely states that many queries are required to solve our problem, we should allow for `powerful' queries in order to strengthen our main result.

\vspace{\baselineskip} \noindent \textsc{Definition:} Fix a setting. A {\it query} is a tuple $\mathsf{q} = (i, \kappa, S, p)$ specifying (i)~a {\it cutter} $i \in N$, (ii)~a knife $\kappa \in \mathbb{K}$, (iii)~a serving $S \in \mathcal{S}$, and (iv)~a target proportion $p \in [0, 1]$. The interpretation is as follows:
\begin{itemize}
\item First, the mediator uses the knives to construct $S$.

\item Second, the mediator asks $i$ to use both his private measure and the knife $\kappa$ to cut~$S$ in order to construct $S' \subseteq S$ with the property that, according to his private measure, the proportion of the value of $S'$ to the value of $S$ is precisely the target proportion $p$. To be explicit, if this private measure is $\mu_i$, then the constructed~$S'$ should satisfy $\mu_i(S') = p \cdot \mu_i(S)$. Furthermore, the mediator provides detailed instructions about how $S'$ should be constructed; see the next section.

\item Third, the mediator asks every agent to use his private measure to appraise both the value of $S$ and the value of $S'$.
\end{itemize}
We let $\mathcal{Q}$ denote the set of queries.

\vspace{\baselineskip} This query definition is our adaptation of the `stronger query definition' of \cite{Cseh-Fleiner2020} to our model.

\hypertarget{Section3.2}{}
\subsection{The cutter's serving}

For each query, the cutter is able to construct a serving that satisfies the mediator's requirements because his preferences are represented by a kitchen measure. Moreover, if there are several such servings, then it does not matter which the cutter selects: in all cases the mediator receives the same information because he knows which servings are null. It is therefore without loss of generality to assert that the cutter is given specific instructions for how to break ties, and doing so simplifies the presentation.

\vspace{\baselineskip} \noindent \textsc{Definition:} Fix a setting. For each query $\mathsf{q} = (i, \kappa, S, p)$ and each $\mu_i \in \mathbb{M}$, we define the {\it $(\mathsf{q}|\mu_i)$-threshold} by $\tau(\mathsf{q} | \mu_i) \equiv \inf \{t \in [0, 1] | \mu_i(S \cap \kappa_t) \geq p \cdot \mu_i(S) \}$ and we define the {\it $(\mathsf{q}|\mu_i)$-serving} by $S_{\mathsf{q} | \mu_i} \equiv S \cap \kappa_{\tau(\mathsf{q} | \mu_i)}$. The interpretation is that when the mediator selects query $\mathsf{q}$ and the cutter $i$ has preferences represented by $\mu_i$, the output serving is $S_{\mathsf{q} | \mu_i}$.

\vspace{\baselineskip} The intended interpretation is that in order to construct $S_{\mathsf{q} | \mu_i}$, the cutter $i$ glides $\kappa$ over $S$, starting at time zero, until the first moment that $S \cap \kappa_t$ is worth at least the target value, at which point he cuts. By the following lemma, the intended interpretation is valid.

\hypertarget{Lemma1}{}
\vspace{\baselineskip} \noindent \textsc{Lemma 1:} Fix a setting. For each query $\mathsf{q} = (i, \kappa, S, p)$ and each $\mu \in \mathbb{D}$, the constructed serving satisfies the mediator's request: $\mu_i(S_{\mathsf{q} | \mu_i}) = p \cdot \mu_i(S)$.

\vspace{\baselineskip} \noindent \textsc{Proof:} Assume the hypotheses. Since $\mu_i$ is a kitchen measure, there is $t \in [0, 1]$ such that $\mu_i(S \cap \kappa_t) = p \cdot \mu_i(S)$. Then $\tau(\mathsf{q} | \mu_i) \leq t$, so $S_{\mathsf{q} | \mu_i} = S \cap \kappa_{\tau(\mathsf{q} | \mu_i)} \subseteq S \cap \kappa_t$, so $\mu_i(S_{\mathsf{q} | \mu_i}) \leq \mu_i(S \cap \kappa_t) = p \cdot \mu_i(S)$.

Assume, by way of contradiction, that $\mu_i(S_{\mathsf{q} | \mu_i}) < p \cdot \mu_i(S)$. Then there is $p' \in [0, 1]$ such that $\mu_i(S_{\mathsf{q} | \mu_i}) < p' \cdot \mu_i(S) < p \cdot \mu_i(S)$, so as $\mu_i \in \mathbb{M}$ there is $t' \in [0, 1]$ such that $\mu_i(S \cap \kappa_{t'}) = p' \cdot \mu_i(S) > \mu_i(S_{\mathsf{q} | \mu_i})$, so $t' > \tau(\mathsf{q} | \mu_i)$. But then by definition of $\tau(\mathsf{q} | \mu_i)$ we have $\mu_i(S \cap \kappa_{t'}) \geq p \cdot \mu_i(S)$, contradicting $\mu_i(S \cap \kappa_{t'}) = p' \cdot \mu_i(S) < p \cdot \mu_i(S)$. Altogether, then, $\mu_i(S_{\mathsf{q} | \mu_i}) = p \cdot \mu_i(S)$, as desired.~$\blacksquare$

\hypertarget{Section3.3}{}
\subsection{Records, restrictions, and responses}

Initially, the mediator knows only that the measure profile belongs to the unrestricted domain $\mathbb{D}$, but after selecting a query and receiving the agents' response, the mediator learns that the measure profile belongs to a restricted domain $\mathbb{D}' \subseteq \mathbb{D}$. More generally, from a list of queries and their responses, the mediator learns a restricted domain. We begin with the general definitions and then define the response to an individual query as a special case.

\vspace{\baselineskip} \noindent \textsc{Definition:} Fix a setting. For each collection of servings $\mathcal{S}' \subseteq \mathcal{S}$, an {\it $\mathcal{S}'$-appraisal} is a function $\alpha_0: \mathcal{S}' \to [0, 1]$. A {\it (finite) record} is a pair $(\mathcal{S}', \alpha')$ such that (i)~$\mathcal{S}' \subseteq \mathcal{S}$ with $|\mathcal{S}'| \in \mathbb{N}$, and (ii)~$\alpha' = (\alpha'_i)_{i \in N}$ is a profile of $\mathcal{S}'$-appraisals. In this case, we define the associated restricted domain to be the collection of kitchen measure profiles that extend the appraisal profile, $\mathbb{D}(\mathcal{S}', \alpha') \equiv \{ \mu \in \mathbb{D} | \text{for each } i \in N,  \alpha'_i = \mu_i \restriction_{\mathcal{S}'}\}$, and we say that $(\mathcal{S}', \alpha')$ is {\it valid} if $\mathbb{D}(\mathcal{S}', \alpha') \neq \emptyset$.\footnote{We are mostly interested in valid records, but for some arguments it is convenient to work with records that are not valid because they assign zero to servings that are not slivers.}

\vspace{\baselineskip} \noindent \textsc{Definition:} Fix a setting. Given a pair of records $(\mathcal{S}', \alpha')$ and $(\mathcal{S}'', \alpha'')$, we say that $(\mathcal{S}'', \alpha'')$ is {\it at least as informative as} $(\mathcal{S}', \alpha')$ if $\mathbb{D}(\mathcal{S}'', \alpha'') \subseteq \mathbb{D}(\mathcal{S}', \alpha')$.

\vspace{\baselineskip} Initially, the mediator does not know how any agent appraises any non-null serving, but at any later stage the mediator will have some query responses that allow him to fill in some of the blanks. The resulting record is a collection of servings together with their appraisal from all agents, and this record in turn functions as a domain restriction: the mediator is sure that the measure profile extends the appraisal profile. A special case is the record consisting of a response to a single query.

\vspace{\baselineskip} \noindent \textsc{Definition:} Fix a setting. For each query $\mathsf{q} = (i, \kappa, S, p)$ and each $\mu \in \mathbb{D}$, the {\it $(\mathsf{q}|\mu)$-response}, $\mathsf{r}(\mathsf{q} | \mu)$, is the record $(\mathcal{S}', \alpha')$ such that (i)~$\mathcal{S}' = \{S, S_{\mathsf{q}|\mu_i}\}$, and (ii)~for each $j \in N$, $\alpha'_j = \mu_j \restriction_{\mathcal{S}'}$. In this case, the {\it $(\mathsf{q}|\mu)$-domain}, $\mathbb{D}(\mathsf{q}|\mu)$, is $\mathbb{D}(\mathsf{r}(\mathsf{q} | \mu))$.

\vspace{\baselineskip} This definition fits our earlier discussion of queries. Indeed, suppose the mediator selects $\mathsf{q}$ and the measure profile is $\mu$. Then first, the mediator constructs $S$; second, the cutter constructs $S_{\mathsf{q}|\mu_i}$; and third, every agent appraises both $S$ and $S_{\mathsf{q}|\mu_i}$. As a result, the mediator learns that the measure profile belongs to $\mathbb{D}(\mathsf{q}|\mu) \subseteq \mathbb{D}$.

\hypertarget{Section3.4}{}
\subsection{Protocols}

The mediator must communicate with the agents until he can identify an allocation that is surely satisfactory, in the sense that it respects the entitlements, given what he has learned. Communication is costly, however, and thus the mediator seeks to select a (communication) protocol that keeps costs down, subject to the constraint that it always identifies a satisfactory allocation.

We formalize a protocol as a pure strategy for the mediator in the following game against nature: nature opens with an unobserved selection of the measure profile, and thereafter the mediator iteratively selects queries and receives responses until he ends the game by selecting an allocation. We begin by formalizing our game against nature, minimizing the notation that we introduce in anticipation of our proof's requirements.

\vspace{\baselineskip} \noindent \textsc{Definition:} Fix a setting. For each $e \in E$, the {\it division game given $e$} is the one-player game against nature defined as follows.
\begin{itemize}
\item The only player with payoffs is the {\it mediator}, and {\it nature} also plays. There are also histories partially ordered by precedence, active players and actions associated with non-terminal histories, information sets, and payoffs. The precedence order is such that each history and action at that history together determine an {\it immediate successor}, or an earliest history that follows. We define these remaining ingredients below, defining precedence entirely using immediate successors.

\item The initial history is non-terminal, its active player is nature, and its set of available actions is~$\mathbb{M}^N$. Nature's choice of measure profile is not observed by the mediator. After each choice of nature, the immediate successor is non-terminal.

\item At each non-terminal history other than the initial history, the active player is the mediator and the set of available actions is $\mathcal{Q} \cup \mathcal{X}$. We denote the set of histories where the mediator is the active player by $H$. If the mediator selects a query, then the immediate successor is non-terminal. If the mediator selects an allocation, then the immediate successor is terminal.

\item Each non-initial history $h$ is identified with the finite sequence of actions that leads to $h$ from the initial history: there is a {\it (query) count} $c \in \mathbb{N}_0$ such that $h = (\mu, \mathsf{q}_1, \mathsf{q}_2, ..., \mathsf{q}_c)$. In this case, the {\it chronicle at $h$} is the list of queries and their responses, $(\mathsf{q}_1, \mathsf{r}(\mathsf{q}_1|\mu), \mathsf{q}_2, \mathsf{r}(\mathsf{q}_2|\mu), ..., \mathsf{q}_c, \mathsf{r}(\mathsf{q}_c|\mu))$. Moreover, the {\it $h$-domain} is $\mathbb{D}(h) \equiv \cap_{t \in \{1, 2, ..., c\}} \mathbb{D}(\mathsf{r}(\mathsf{q}_t|\mu))$, where by convention $c=0$ implies $\mathbb{D}(h) = \mathbb{D}$; this is nonempty as $\mu \in \mathbb{D}(h)$. Finally, two histories share an {\it information set} if and only if (i)~the mediator is the active player at both histories, and (ii)~the two histories have the same chronicle. Observe that if $h$ and $h'$ share an information set, then $\mathbb{D}(h) = \mathbb{D}(h')$.

\item A play is a maximal chain of histories. If a play ends at a terminal history with a query count of $c$, if the measure profile selected by nature is $\mu$, and if the allocation selected by the mediator is $(e|\mu)$-proportional, then the mediator's {\it payoff} is $-c$; otherwise it is $-\infty$.
\end{itemize}
In this case, when we delete the payoffs, the result is the {\it division game form}. Note that our main result about communication costs does not directly involve these payoffs, but rather the related costs that we introduce shortly; thus at this point the payoffs can be taken as suggestive.

\vspace{\baselineskip} A protocol is a plan for the mediator in the division game form that is contingent only upon the available information.

\vspace{\baselineskip} \noindent \textsc{Definition:} Fix a setting. A {\it protocol} is a pure strategy for the mediator in the division game form. Equivalently, a protocol is a function $\pi: H \to \mathcal{Q} \cup \mathcal{X}$ such that for each pair $h, h' \in H$, if $h$ and $h'$ share an information set, then $\pi(h) = \pi(h')$. In this case, for each $\mu \in \mathbb{D}$, we define $\mathsf{cost}(\pi|\mu) \in \mathbb{N}_0 \cup \{\infty\}$ and $\mathsf{out}(\pi|\mu) \in \mathcal{X} \cup \{\mathsf{error}\}$ as follows.
\begin{itemize}
\item If the play determined by $\mu$ and $\pi$ is infinite, then we define $\mathsf{cost}(\pi|\mu) \equiv \infty$ and $\mathsf{out}(\pi|\mu) \equiv \mathsf{error}$.

\item If the play determined by $\mu$ and $\pi$ is finite, then let $c$ denote the query count of its terminal history and let $X$ denote its final action. In this case, we define $\mathsf{cost}(\pi|\mu) \equiv c$ and $\mathsf{out}(\pi|\mu) \equiv X$.
\end{itemize}
We let $\Pi$ denote the set of protocols. Observe that for each $e \in E$, the mediator's payoff in the division game given $e$ for the play determined by $\mu$ and $\pi$ is (i)~$-\infty$, if either $\mathsf{out}(\pi|\mu) = \mathsf{error}$ or $\mathsf{out}(\pi|\mu)$ is not $(e|\mu)$-proportional, and (ii)~$-\mathsf{cost}(\pi|\mu)$, otherwise.

\vspace{\baselineskip} We assess each protocol in the context of a given entitlement profile $e$, but instead of doing so using payoffs directly, we separately assess (i)~whether or not the protocol always constructs $e$-proportional allocations, and (ii)~the worst-case communication cost measured in queries. We begin with the former.

\vspace{\baselineskip} \noindent \textsc{Definition:} Fix a setting. For each $e \in E$, a protocol is {\it $e$-proportional} if for each $\mu \in \mathbb{M}^N$, $\mathsf{out}(\pi|\mu) \in \mathcal{X}$ and $\mathsf{out}(\pi|\mu)$ is $(e|\mu)$-proportional. We let $\Pi_e \subseteq \Pi$ denote the set of $e$-proportional protocols.

\vspace{\baselineskip} The set of $e$-proportional protocols is known to be nonempty (\citealp{Barbanel1995}; \citealp{Shishido-Zeng1999}; \citealp{Cseh-Fleiner2020}; \citealp{Janko-Joo2022}), and we are interested in comparing these protocols according to the following performance metric.

\vspace{\baselineskip} \noindent \textsc{Definition:} Fix a setting and a protocol. The {\it worst-case cost of $\pi$} is
\begin{align*}
\mathsf{cost}(\pi) \equiv \sup_{\mu \in \mathbb{D}} \mathsf{cost}(\pi|\mu).
\end{align*}
If $\mathsf{cost}(\pi) = \infty$, then we say that $\pi$ is {\it unbounded}; otherwise we say that $\pi$ is {\it bounded}.

\vspace{\baselineskip} Altogether, then, the mediator's problem is given by a setting and an associated entitlement profile $e$, and the objective is to minimize the worst-case cost among the $e$-proportional protocols.

\vspace{\baselineskip} \noindent \textsc{Definition:} Fix a setting. For each $e \in E$, the {\it optimal worst-case cost of an $e$-proportional protocol} is
\begin{align*}
\mathsf{cost}(e) \equiv \min \{ c \in \mathbb{N}_0 \cup \{\infty\} | \text{ there is } \pi \in \Pi_e \text{ such that } \mathsf{cost}(\pi) = c \}.
\end{align*}
Because the set of $e$-proportional protocols is nonempty (\citealp{Barbanel1995}; \citealp{Shishido-Zeng1999}; \citealp{Cseh-Fleiner2020}; \citealp{Janko-Joo2022}), this is well-defined.

\hypertarget{Section4}{}
\section{Communication costs}

\hypertarget{Section4.1}{}
\subsection{The mediator's problem}

In their important contribution, \cite{Cseh-Fleiner2020} associate each problem that the mediator might face with both a lower bound and an upper bound on its cost. These bounds and our analysis together involve three indices for entitlement profiles: clonage, precision, and fineness. The most important index is clonage, which we ultimately use to present both our lower bound and the Cseh-Fleiner upper bound. Our lower bound analysis involves precision, while the Cseh-Fleiner lower bound involves fineness, and crucially clonage can only be bounded as a function of the former.

If we restrict attention to entitlement profiles with rational and positive entitlements, then the three indices admit simple verbal descriptions.
\begin{itemize}
\item The {\it clonage} is the minimum number of clones required to form an equal-entitlement economy, provided we must replace each agent~$i$ with $n_i \in \mathbb{N}$ clones who equally split $e_i$. This is the least common multiple of the denominators of the entitlements' reduced fractions.

\item The {\it precision} is the maximum of the denominators of the entitlements' reduced fractions.

\item The {\it fineness} is the smallest denominator of a positive fraction that is at most the smallest positive entitlement.
\end{itemize}
The formal definitions, which apply to all entitlement profiles, are as follows.

\vspace{\baselineskip} \noindent \textsc{Definition:} Fix a setting. For each $m \in \mathbb{N}$, define the set of {\it $m$-ruler markings} by $\mathcal{M}_m \equiv \{ \frac{a}{m} | a \in \{0, 1, ..., m\} \}$ and the set of {\it positive $m$-ruler markings} by $\mathcal{M}^+_m \equiv \mathcal{M}_m \backslash \{0\}$. For each $e \in E$, we define the following.
\begin{itemize}
\item The {\it clonage of $e$} is $\mathsf{C}(e) \equiv \min (\{m \in \mathbb{N} | \text{ for each } i \in N, e_i \in \mathcal{M}_m \} \cup \{\infty\})$.

\item For each $i \in N$, $\mathsf{P}_i(e) \equiv \min (\{m \in \mathbb{N} | e_i \in \mathcal{M}_m \} \cup \{\infty\})$. The {\it precision of $e$} is $\mathsf{P}(e) \equiv \max_{i \in N} \mathsf{P}_i(e)$.

\item For each $i \in N$, $\mathsf{F}_i(e) \equiv \min \{m \in \mathbb{N} | e_i \in \{0\} \cup [\min \mathcal{M}^+_m, 1] \}$. The {\it fineness of~$e$} is $\mathsf{F}(e) \equiv \max_{i \in N} \mathsf{F}_i(e)$.
\end{itemize}
When $\mathsf{C}$, $\mathsf{P}$, and $\mathsf{F}$ are viewed as functions from $E$ to $\mathbb{N} \cup \{\infty\}$, we refer to them as the {\it clonage index}, the {\it precision index}, and the {\it fineness index}, respectively.

\vspace{\baselineskip} The Cseh-Fleiner upper bound involves clonage.

\hypertarget{TheoremCF1}{}
\vspace{\baselineskip} \noindent \textsc{Theorem CF1:}\footnote{This is Theorem~5.5 of \cite{Cseh-Fleiner2020}.} Fix a setting. For each $e \in E$ and each clonage level $\mathsf{c} \in \mathbb{N}$, $\mathsf{C}(e) = \mathsf{c}$ implies $\mathsf{cost}(e) \leq 2(n-1) \lceil \log_2 \mathsf{c} \rceil$.

\vspace{\baselineskip} By contrast, at the entitlement profile level, the Cseh-Fleiner lower bound involves fineness.

\hypertarget{TheoremCF2}{}
\vspace{\baselineskip} \noindent \textsc{Theorem CF2:}\footnote{This is a variant of Theorem~6.3 in \cite{Cseh-Fleiner2020} involving strict inequalities instead of weak inequalities; a minor modification of the original proof suffices for this version.} Fix a setting. For each $e \in E$ and each fineness level $\mathsf{f} \in \mathbb{N}$, $\mathsf{F}(e) > \mathsf{f}$ implies $\mathsf{cost}(e) > \log_3 \mathsf{f}$.

\vspace{\baselineskip} We sketch the intuition for this lower bound, which remarkably applies even when (i)~there are two agents, (ii)~we know one agent's measure, and (iii)~after each query we learn the other agent's appraisal of the new partition generated by the old partition and the query. Indeed, suppose we know $\mu_1$, and suppose the entitlement profile has fineness greater than $\mathsf{f}$ because $\frac{1}{\mathsf{f}} > e_2 > 0$. Any query and response together divide each cell of the old partition into at most three new sub-cells, and \cite{Cseh-Fleiner2020} prove that it is always possible for $2$ to respond that all of each old cell's value---except, perhaps, a negligible amount\footnote{This modification is needed in our model because the mediator knows the set of null servings precisely.}---resides entirely in its new sub-cells that $1$ values most. In this case, after $q$ queries, a non-negligible cell for $2$ is worth at least $\frac{1}{3^q}$ to $1$, we must give such a cell $S$ to $2$, and in order for $1$ to agree we must have $\frac{1}{3^q} \leq \mu_1(S) \leq 1 - e_1 = e_2 < \frac{1}{\mathsf{f}}$; thus the number of queries $q$ must be great enough given the fineness $\mathsf{f}$.

Our first proposition provides a lower bound at the entitlement profile level using precision instead of fineness.

\hypertarget{Proposition1}{}
\vspace{\baselineskip} \noindent \textsc{Proposition 1:} Fix a setting. For each $e \in E$ and each precision level $\mathsf{p} \in \mathbb{N}$, $\mathsf{P}(e) > \mathsf{p}$ implies $\mathsf{cost}(e) > \lfloor \log_2 \log_2 2 \mathsf{p} \rfloor$.

\vspace{\baselineskip} The proof spans \hyperlink{Appendix1}{Appendix~1}, \hyperlink{Appendix2}{Appendix~2}, and \hyperlink{Appendix3}{Appendix~3}. At a high level, the first appendix establishes our main two-agent lemma for queries that select a cell of the current partition (\hyperlink{Lemma3}{Lemma~3}), the second appendix establishes our generalization of \hyperlink{Lemma3}{Lemma~3} to all queries (\hyperlink{Lemma4}{Lemma~4}), and the third appendix applies \hyperlink{Lemma4}{Lemma~4} to establish the lower bound for any number of agents (\hyperlink{Proposition1}{Proposition~1}). The key idea is that by cloning the cake instead of the agents, we can index the deficiency of a collection of responses. We discuss this proof in more detail in \hyperlink{Section4.3}{Section~4.3}. By the following result of \cite{Lohr2012}, \hyperlink{Proposition1}{Proposition~1} is, in a suitable sense, tight.

\hypertarget{TheoremL}{}
\vspace{\baselineskip} \noindent \textsc{Theorem L \citep{Lohr2012}:} Fix a setting with two agents. There is an increasing sequence of precision levels such that for each precision level $\mathsf{p}$ in the sequence, there is $e \in E$ such that $\mathsf{P}(e) = \mathsf{p}$ and $\mathsf{cost}(e) \leq (\log_2 \log_2 \mathsf{p})+1$.

\vspace{\baselineskip} The original presentation of the Cseh-Fleiner lower bound involves clonage, and we include such a statement as part of \hyperlink{CorollaryCF}{Corollary~CF} in \hyperlink{Section4.2}{Section~4.2}. Crucially, however, that statement is at the setting level. At the entitlement profile level, a lower bound involving fineness cannot be translated to a lower bound involving clonage, yet a lower bound involving precision can, and this is a consequence of the relationships between the three indices described by the following proposition.

\hypertarget{Proposition2}{}
\vspace{\baselineskip} \noindent \textsc{Proposition 2:} The three indices can be compared as follows.
\begin{itemize}
\item For each setting and each $e \in E$, we have $\mathsf{C}(e) \geq \mathsf{P}(e) \geq \mathsf{F}(e)$.

\item There are a setting and an $e \in E$ such that $\mathsf{C}(e) > \mathsf{P}(e)$.

\item For each setting with $n \geq 2$ and each $e \in E$, $\mathsf{C}(e) = \infty$ implies $\mathsf{P}(e) = \infty$ and $\mathsf{C}(e) \in \mathbb{N}$ implies $\mathsf{P}(e)^{n-1} \geq \mathsf{C}(e)$.

\item For each setting with $n \geq 2$, there is no $f: \mathbb{N} \to \mathbb{N}$ such that for each $e \in E$, $\mathsf{P}(e) \in \mathbb{N}$ implies $f(\mathsf{F}(e)) \geq \mathsf{P}(e)$.
\end{itemize}

\vspace{\baselineskip} \noindent \textsc{Proof:} We begin with the first two items. The first item is immediate from definitions; we omit the proof. For the second item, an example is any setting with $N = \{1, 2, 3, 4\}$, $e_1 = \frac{1}{2}$, $e_2 = \frac{1}{3}$, $e_3 = \frac{1}{10}$, and $e_4 = \frac{1}{15}$; in this case $\mathsf{C}(e) = 30$ and $\mathsf{P}(e) = 15$.

For the third item, fix a setting with $n \geq 2$ and let $e \in E$. If $\mathsf{C}(e) = \infty$, then there is an irrational entitlement and thus $\mathsf{P}(e) = \infty$, as desired. If $\mathsf{C}(e) \in \mathbb{N}$, then define $\mathsf{c} \equiv \mathsf{C}(e)$ and define $\mathsf{p} \equiv \mathsf{P}(e)$. Moreover, for each $i \in N$, define $b_i \equiv \mathsf{P}_i(e)$. Re-index the agents using $N = \{1, 2, ..., n\}$ such that $b_1 \geq b_2 \geq ... \geq b_n$, then define $b \equiv b_1 \cdot b_2 \cdot ... \cdot b_{n-1}$. For each $i \in N \backslash \{n\}$, there is $a_i \in \{0, 1, ..., b_i\}$ such that $e_i = \frac{a_i}{b_i}$, so there is $a'_i \in \{0, 1, ..., b\}$ such that $e_i = \frac{a'_i}{b}$, so $e_i \in \mathcal{M}_b$. Moreover, $e_n = 1 - \sum_{i \in N \backslash \{n\}} e_i = \frac{b - \sum_{i \in N \backslash \{n\}} a'_i}{b}$, so $e_n \in \mathcal{M}_b$. Altogether, then, $\mathsf{c} \leq b \leq b_1^{n-1} = \mathsf{p}^{n-1}$, as desired.

For the final item, assume by way of contradiction we have such a function $f$. For each $m \in \{3, 4, ...\}$, let $e^m \in E$ be such that one agent has entitlement $\frac{1}{2} - \frac{1}{2m+1}$, another agent has entitlement $\frac{1}{2} + \frac{1}{2m+1}$, and all other agents have entitlement zero. Then for each $m \in \{3, 4, ...\}$, we have $\mathsf{F}(e^m) = 3$ and $\mathsf{P}(e^m) = 4m+2$. But then there is $m \in \{3, 4, ...\}$ such that $\mathsf{P}(e^m) > f(3) = f(\mathsf{F}(e^m))$, contradicting that $ f(\mathsf{F}(e^m)) \geq \mathsf{P}(e^m)$.~$\blacksquare$

\vspace{\baselineskip} Using \hyperlink{Proposition2}{Proposition~2}, we can translate \hyperlink{Proposition1}{Proposition~1} from a lower bound involving precision to one involving clonage.

\hypertarget{Theorem1}{}
\vspace{\baselineskip} \noindent \textsc{Theorem 1 (Main result):} Fix a setting with at least two agents. For each clonage level $\mathsf{c} \in \mathbb{N}$, $\mathsf{C}(e) > \mathsf{c}$ implies $\mathsf{cost}(e) > \lfloor \log_2 \log_2 2 \mathsf{c}^{\frac{1}{n-1}} \rfloor$.

\vspace{\baselineskip} If there is one agent, of course, then for the unique member of $E$ we have $\mathsf{cost}(e) = 0$. We conclude this section with our titular corollary about the irrationally entitled.

\hypertarget{Corollary1}{}
\vspace{\baselineskip} \noindent \textsc{Corollary 1:} Fix a setting and an entitlement profile $e$. If there is an agent whose entitlement is irrational, then no $e$-proportional protocol is bounded.

\hypertarget{Section4.2}{}
\subsection{The meta-problem}

In this section, we use \hyperlink{Theorem1}{Theorem~1} to analyze how cost varies as the problem varies. To do so, we consider the meta-problem of selecting a protocol for each problem, also known as the algorithm selection problem \citep{Rice1976}. The distinction between the mediator's problem and the meta-problem is worth emphasizing.
\begin{itemize}
\item A {\it problem} is given by a setting and an entitlement profile $e$. In this case, an {\it instance} is a kitchen measure profile $\mu$, whose associated {\it solutions} are the $(e|\mu)$-proportional allocations, and a {\it protocol} maps each instance to one of its solutions.

\item A {\it meta-problem} is given by a setting. In this case, a {\it meta-instance} is an associated problem given by an entitlement profile $e$, whose associated {\it meta-solutions} are the $e$-proportional protocols, and (iii)~a {\it meta-protocol} maps each meta-instance to one of its meta-solutions.
\end{itemize}
We note that our methodology for analyzing the meta-problem, which is standard in the cake division literature, involves parameterized complexity theory (\citealp{Downey-Fellows1999}; \citealp{Flum-Grohe2006}). In particular, in the terminology of \cite{Flum-Grohe2006}, a parameterization of a problem is a function that associates each instance with a parameter. We take the clonage index as our parameterization of the meta-problem, while other papers on cake division take the number of agents as the parameterization.

For the mediator's problem, which protocols are optimal depends on the performance metric, and we focused on a single metric: the worst-case communication cost measured in queries. For the meta-problem, the optimal meta-protocols are fixed as those that associate each problem with an optimal protocol, and we measure how their performance varies with clonage using two metrics.

\vspace{\baselineskip} \noindent \textsc{Definition:} Fix a setting. Let $\mathsf{infcost}: \mathbb{N} \to \mathbb{R}_+ \cup \{\infty\}$ and $\mathsf{supcost}: \mathbb{N} \to \mathbb{R}_+ \cup \{\infty\}$ be defined such that for each clonage level $\mathsf{c} \in \mathbb{N}$,
\begin{align*}
\mathsf{infcost}(\mathsf{c}) &\equiv \inf \{\mathsf{cost}(e) | e \in E \text{ and } \mathsf{C}(e) = \mathsf{c} \}, \text{ and}
\\\mathsf{supcost}(\mathsf{c}) &\equiv \sup \{\mathsf{cost}(e) | e \in E \text{ and } \mathsf{C}(e) = \mathsf{c} \}.
\end{align*}
Recall that for each $f: \mathbb{N} \to \mathbb{R}_+ \cup \{\infty\}$ and each $g: \mathbb{N} \to \mathbb{R}_+$, we use $g$ to express the asymptotic order of $f$ by writing
\begin{itemize}
\item {\it $f(x) = \mathcal{O}(g(x))$} if and only if there are $x^* \in \mathbb{N}$ and $M^* \in \mathbb{R}_+$ such that for each $x \in \{x^*, x^*+1, ...\}$, $f(x) \leq M^* \cdot g(x)$, and

\item {\it $f(x) = \Omega(g(x))$} if and only if there are $x^* \in \mathbb{N}$ and $M^* \in \mathbb{R}_+$ such that for each $x \in \{x^*, x^*+1, ...\}$, $f(x) \geq M^* \cdot g(x)$.
\end{itemize}
We are interested in expressing the asymptotic order of $\mathsf{infcost}$ and $\mathsf{supcost}$.

\vspace{\baselineskip} \hyperlink{TheoremCF1}{Theorem~CF1} and \hyperlink{TheoremCF2}{Theorem~CF2} together bound $\mathsf{supcost}$ both above and below, and our results do not improve either bound.

\hypertarget{CorollaryCF}{}
\vspace{\baselineskip} \noindent \textsc{Corollary~CF \citep{Cseh-Fleiner2020}:} Fix a setting. We have $\mathsf{supcost}(\mathsf{c}) = \mathcal{O}(\log_2 \mathsf{c})$ and $\mathsf{supcost}(\mathsf{c}) = \Omega(\log_3 \mathsf{c}) =  \Omega(\log_2 \mathsf{c})$.

\vspace{\baselineskip} Instead, we provide a lower bound on $\mathsf{infcost}$.

\hypertarget{Corollary2}{}
\vspace{\baselineskip} \noindent \textsc{Corollary~2:} Fix a setting. We have $\mathsf{infcost}(\mathsf{c}) = \Omega(\log_2 \log_2 \mathsf{c})$.

\vspace{\baselineskip} Together, \hyperlink{CorollaryCF}{Corollary~CF} and \hyperlink{Corollary2}{Corollary~2} allow us to formalize our key insight that clonage causes communication costs: a sufficient increase in clonage guarantees that costs increase.

\hypertarget{Corollary3}{}
\vspace{\baselineskip} \noindent \textsc{Corollary~3:} Fix a setting. For each clonage level $c \in \mathbb{N}$, there is clonage level $c' \in \mathbb{N}$ with $c' > c$ such that for each pair $e, e' \in E$, $\mathsf{C}(e) \leq c$ and $\mathsf{C}(e') \geq c'$ implies $\mathsf{cost}(e) < \mathsf{cost}(e')$.

\hypertarget{Section4.3}{}
\subsection{Proof discussion}

Our main result follows from \hyperlink{Proposition1}{Proposition~1} and \hyperlink{Proposition2}{Proposition~2}, and we have already proven the latter. In this section, we discuss the proof of the former. We begin by introducing the concepts needed to state our key lemma, \hyperlink{Lemma3}{Lemma~3}.

To begin, sometimes two records are equally informative in the sense that they are associated with the same restricted domain. It is sometimes convenient to use a record with a small collection of servings, and it is sometimes convenient to use a record with a large number of servings. In our analysis, we restrict attention to the case that the small collection consists of the cells of a partition and the large collection is the associated subalgebra.

\vspace{\baselineskip} \noindent \textsc{Definition:} Fix a setting. A (finite) record $(\mathcal{S}', \alpha')$ is a {\it (finite) partition record} if $\mathcal{S}'$ is a finite partition. In this case, we often emphasize that the record is a partition record by writing $(\mathcal{P}, \alpha)$. We let $\mathcal{A}_\mathcal{P} \subseteq \mathcal{S}$ denote the subalgebra of $\mathcal{S}$ generated by $\mathcal{P}$, and for each $i \in N$, we let $\alpha^\uparrow_i$ denote the {\it subalgebra extension of $\alpha_i$ (to $\mathcal{A}_\mathcal{P}$)}: the function $\alpha^\uparrow_i: \mathcal{A}_\mathcal{P} \to [0, 1]$ such that for each $S \in \mathcal{A}_\mathcal{P}$ we have $\alpha^\uparrow_i(S) = \sum_{\{S' \in \mathcal{P} | S' \subseteq S\}} \alpha_i(S')$.

\vspace{\baselineskip} We can always use a partition record as a concise upper bound on the information available to the mediator, in which case we conveniently have an equivalent subalgebra record.

\vspace{\baselineskip} \noindent \textsc{Observation:} Fix a setting and let $(\mathcal{S}', \alpha')$ be a (finite) record. For each record $(\mathcal{P}, \alpha)$ such that (i)~$\mathcal{P}$ is the collection of atoms in the subalgebra of $\mathcal{S}$ generated by $\mathcal{S}'$, and (ii)~for each $i \in N$ and each $S' \in \mathcal{S}'$, $\alpha^\uparrow_i(S') = \alpha'_i(S')$, we have that $(\mathcal{P}, \alpha)$ is a partition record that is at least as informative as $(\mathcal{S}', \alpha')$.

\vspace{\baselineskip} \noindent \textsc{Observation:}\footnote{We caution that in general, if we are given a (finite) record $(\mathcal{S}', \alpha')$ and seek an equally informative record with more entries, then by iteratively adding uninformative new entries for complements and disjoint unions, we reach a record for the {\it Dynkin system} generated by $\mathcal{S}'$, and a record for the {\it subalgebra} generated by $\mathcal{S}'$ may be strictly more informative. Consider, for example, the interval cake and $\mathcal{S}' = \{(\frac{1}{5}, \frac{3}{5}], (\frac{2}{5}, \frac{4}{5}]\}$: an associated appraisal generally does not allow us to deduce the value of $(\frac{2}{5}, \frac{3}{5}]$.} Fix a setting. For each (finite) partition record $(\mathcal{P}, \alpha)$, the record $(\mathcal{A}_\mathcal{P}, \alpha^\uparrow)$ is equally informative: $\mathbb{D}(\mathcal{P}, \alpha) = \mathbb{D}(\mathcal{A}_\mathcal{P}, \alpha^\uparrow)$.

\vspace{\baselineskip} Our key idea is indexing the difficulty that the mediator faces at an information set using the number of cake clones that he requires to solve the problem with no further queries. For our purposes, it suffices to formalize this difficulty index only for two-agent settings and partition records. To avoid confusion with agent clones, we refer to copies of the cake as {\it replicas}, and we refer to the difficulty that the mediator faces as the partition record's {\it deficiency level}.

\vspace{\baselineskip} \noindent \textsc{Definition:} Fix a setting with agents in $N = \{1, 2\}$ and an entitlement profile $e \in E$. For each partition record $(\mathcal{P}, \alpha)$ and each {\it level} $\ell \in \mathbb{N}$, we define the following.
\begin{itemize}
\item For each $r \in \mathbb{N}$, an {\it $(r|\mathcal{P})$-hyperserving} is a member of $\mathcal{S}_{r|\mathcal{P}} \equiv (\mathcal{A}_\mathcal{P})^r$. The suggested interpretation is that the cake is replaced by $r$ identical replicas, with the property that each serving in a replica is worth $\frac{1}{r}$ of the associated serving in the original cake, and the hyperserving $(S_1, S_2, ..., S_r)$ consists of a serving from each replica that is a union of partition cells.

\item An {\it $(\ell| \mathcal{P})$-hyperallocation} is a member of $\mathcal{X}_{\ell|\mathcal{P}} \equiv \cup_{r \in \{1, 2, ..., \ell\}} \mathcal{S}_{r|\mathcal{P}}$. For each $X \in \mathcal{X}_{\ell|\mathcal{P}}$, we define the {\it number of replicas for $X$}, $r(X)$, to be the unique $r \in \{1, 2, ..., \ell\}$ such that $X \in \mathcal{S}_{r|\mathcal{P}}$. The interpretation is that the cake is replaced by $r(X)$ replicas, then agent $1$ consumes~$X$ while agent $2$ consumes the rest.

\item For each $i \in N$ we define the {\it hyperallocation extension of $\alpha_i^\uparrow$ (to $\mathcal{X}_{\ell| \mathcal{P}}$)}: the function $\alpha_i^{\uparrow \uparrow}: \mathcal{X}_{\ell| \mathcal{P}} \to [0, 1]$ such that for each $X \in \mathcal{X}_{\ell| \mathcal{P}}$ we have $\alpha^{\uparrow\uparrow}_i(X) \equiv \sum_{j = 1}^{r(X)} \frac{\alpha^\uparrow_i(X_j)}{r(X)}$.

\item We say that $(\mathcal{P}, \alpha)$ is {\it $\ell$-deficient} if for each $X \in \mathcal{X}_{\ell|\mathcal{P}}$, either $\alpha^{\uparrow\uparrow}_1(X) < e_1$ or $\alpha^{\uparrow\uparrow}_2(X) > e_1$.
\end{itemize}
Observe that if $(\mathcal{P}, \alpha)$ is $1$-deficient, then no $e$-proportional protocol selects an allocation at an information set whose record is at most as informative as $(\mathcal{P}, \alpha)$.

\vspace{\baselineskip} As we use partition records as upper bounds for the information that the mediator currently has, so too do we use {\it ultraresponses} as upper bounds for the information that the mediator acquires.

\vspace{\baselineskip} \noindent \textsc{Definition:} Fix a setting. For each valid partition record $(\mathcal{P}, \alpha)$, each query $\mathsf{q} = (i, \kappa, S, p)$, and each $\mu \in \mathbb{D}(\mathcal{P}, \alpha)$, we define the {\it $(\mathsf{q} | \mu, \mathcal{P}, \alpha)$-ultraresponse}, $\mathsf{r}(\mathsf{q}| \mu, \mathcal{P}, \alpha)$, to be the partition record $(\mathcal{P}^+, \alpha^+)$ such that (i)~$\mathcal{P}^+$ is the collection of atoms in the subalgebra of $\mathcal{S}$ generated by $\mathcal{P} \cup \{S, S_{\mathsf{q}|\mu_i}\}$, and (ii)~for each $j \in N$, $\alpha^+_j = \mu_j\restriction_{\mathcal{P}^+}$. We denote the collection of these ultraresponses by $\mathcal{R}(\mathsf{q}| \mathcal{P}, \alpha) \equiv \{\mathsf{r}(\mathsf{q} | \mu, \mathcal{P}, \alpha) | \mu \in \mathbb{D}(\mathcal{P}, \alpha)\}$.

\vspace{\baselineskip} We want to formalize the idea that if the mediator finds himself in a particularly difficult situation, then for each query he might ask, there is a response that leaves him in a situation that is still rather difficult. Our key lemma establishes this for the special case that the mediator's query involves a cell of the current partition.

\hypertarget{Lemma3}{}
\vspace{\baselineskip} \noindent \textsc{Lemma 3:} Fix a setting with two agents in $N = \{1, 2\}$ and let $e \in E$. For each level~$\ell$, each valid partition record $(\mathcal{P}, \alpha)$ that is $\ell$-deficient, and each $\textsf{q} \in \mathcal{Q}$ such that the serving argument of $\mathsf{q}$ belongs to $\mathcal{P}$, there is $(\mathcal{P}^+, \alpha^+) \in \mathcal{R}(\mathsf{q}|\mathcal{P}, \alpha)$ that is $\lfloor \sqrt{\ell/2} \rfloor$-deficient.

\vspace{\baselineskip} \noindent \textsc{Proof sketch:} The proof (see \hyperlink{Appendix1}{Appendix~1}) is largely constructive. In response to a query about a cell $B$, one of the agents cuts $B$ into two sub-cells $B'$ and $B''$, resulting in $\mathcal{P}^+$, and these details do not matter. All that matters is one number: how much of the value of $B$ does the second agent assign to $B'$? The goal is to find a value $v$ answering this question such that for $\ell^+ \equiv \lfloor \sqrt{\ell/2} \rfloor$, whenever the first agent considers an $(\ell^+|\mathcal{P}^+)$-hyperallocation to be a candidate solution, the second agent rules it out.

Let us say an $(\ell^+|\mathcal{P}^+)$-hyperallocation is $B'$-heavy if the first agent receives more copies of $B'$ than $B''$, $B''$-heavy if the first agent receives more copies of $B''$ than $B'$, and balanced otherwise. The balanced $(\ell^+|\mathcal{P}^+)$-hyperallocations are ruled out because they are equivalent to $(\ell^+|\mathcal{P})$-hyperallocations and $(\mathcal{P}, \alpha)$ is $\ell$-deficient, high values of $v$ rule out $B'$-heavy $(\ell^+|\mathcal{P}^+)$-hyperallocations, and low values of $v$ rule out $B''$-heavy $(\ell^+|\mathcal{P}^+)$-hyperallocations. The problem is therefore only interesting if there are both $B'$-heavy candidates and $B''$-heavy candidates, as in this case we cannot simply rule out the only candidates with an extreme response from the second agent.

The key argument is that in this case, {\it the tightest value threshold above which no $B'$-heavy candidate is accepted} is less than {\it the tightest value threshold below which no $B''$-heavy candidate is accepted}, so that any intermediate value will do; this is Step~7 of the proof. Indeed, otherwise there are a value $v$, a $B'$-heavy candidate $X'$, and a $B''$-heavy candidate $X''$ such that with $v$ the second agent declares both $X'$ and $X''$ acceptable. In this case, however, we show that by appropriately replicating $X'$ and $X''$, we can construct an $(\ell|\mathcal{P}^+)$-hyperallocation that solves the problem and is equivalent to an $(\ell|\mathcal{P})$-hyperallocation, contradicting that $(\mathcal{P}, \alpha)$ is $\ell$-deficient.

\vspace{\baselineskip} From here, \hyperlink{Lemma4}{Lemma~4} extends \hyperlink{Lemma3}{Lemma~3} to arbitrary queries, for which the mediator's selected serving and the cutter's response together divide each existing cell into up to three sub-cells, as in \cite{Cseh-Fleiner2020}. Loosely, the idea is to assign negligible value to sub-cells so that it is almost as though one cell was divided into two sub-cells, so that we can derive the desired conclusion from \hyperlink{Lemma3}{Lemma~3}; see \hyperlink{Appendix2}{Appendix~2} for the details.

We complete the proof of \hyperlink{Proposition1}{Proposition~1} in \hyperlink{Appendix3}{Appendix~3}. First, we use \hyperlink{Lemma4}{Lemma~4} to establish the proposition for the two-agent case. Intuitively, if the entitlement profile has a lot of precision, then the initial information set has a lot of deficiency, so since bad responses can always keep deficiency relatively intact, an unlucky mediator may face a long sequence of bad responses and remain unable to solve the problem. The one-agent case is trivial, so to conclude we handle the case of more than two agents using the two-agent result. Intuitively, we cannot have a protocol whose cost is too low relative to the precision for more than two agents, because then we could imitate that protocol for one of the highest-precision agents in a two-agent setting where the other entitlements are merged, contradicting our two-agent result. We emphasize that this argument relies on precision instead of clonage because precision is defined at the agent level before it is aggregated for the group, but clonage is defined only at the group level; this is why we complete the proof of \hyperlink{Theorem1}{Theorem~1} by separately applying \hyperlink{Proposition2}{Proposition~2}.

\hypertarget{Section5}{}
\section{Further results}

\hypertarget{Section5.1}{}
\subsection{Preference foundation for kitchen measures}

What, exactly, are we assuming about an agent's preferences when we assume they have a kitchen measure representation? We provide an axiomatic answer to this question by modifying a classic result of \cite{Savage1954}, replacing one of Savage's technical axioms with two new axioms involving the knife block.

To begin, an agent's preferences are used to compare servings, and we assume that they satisfy some basic conditions that are necessary to admit a probability measure representation. In particular, we assume that the preferences form a {\it qualitative probability} (\citealp{Bernstein1917}; \citealp{deFinetti1937}; \citealp{Koopman1940}), an object originally interpreted as beliefs about the relative likelihood of uncertain events.

\vspace{\baselineskip} \noindent \textsc{Definition:} Fix a kitchen and let $\succsim$ be a binary relation on $\mathcal{S}$. We say that $\succsim$ satisfies
\begin{itemize}
\item {\it order} if $\succsim$ is complete and transitive;

\item {\it separability} if for each triple $A, B, S \in \mathcal{S}$ such that $A \cap S = B \cap S = \emptyset$, we have $A \succsim B$ if and only if $A \cup S \succsim B \cup S$;

\item {\it monotonicity} if for each pair $A, B \in \mathcal{S}$, $A \subseteq B$ implies $B \succsim A$; and

\item {\it non-degeneracy} if there are $A, B \in \mathcal{S}$ such that $A \succ B$.
\end{itemize}
We say that $\succsim$ is a {\it qualitative probability} if it satisfies all of these axioms.

\vspace{\baselineskip} Of these axioms, the most substantial assumption for cake division is separability. Indeed, this implies that given three equivalent slices, two chocolate and one vanilla, any union of a pair of these slices is equivalent: chocolate-vanilla is no better or worse than chocolate-chocolate.

There are qualitative probabilities without probability measure representations \citep{Kraft-Pratt-Seidenberg1959}. In order to guarantee probability measure representation, \cite{Savage1954} introduced the following technical axioms.

\vspace{\baselineskip} \noindent \textsc{Definition:} Fix a kitchen. We say that a qualitative probability $\succsim$ satisfies
\begin{itemize}
\item {\it fineness} if for each $A \in \mathcal{S}$ such that $A \succ \emptyset$, there are $m \in \mathbb{N}$ and $(B_t)_{t \in \{1, 2, ..., m\}} \in \mathcal{S}^m$ such that (i)~$\{B_1, B_2, ..., B_m\}$ is a partition of $C$, and (ii)~for each $t \in \{1, 2, ..., m\}$, we have $A \succsim B_t$; and

\item {\it tightness} if for each pair $A, B \in \mathcal{S}$, if (i)~for each $A' \in \mathcal{S}$ such that $A' \succ \emptyset$ and $A \cap A' = \emptyset$, we have $A \cup A' \succsim B$, and (ii)~for each $B' \in \mathcal{S}$ such that $B' \succ \emptyset$ and $B \cap B' = \emptyset$, we have $B \cup B' \succsim A$, then we have that $A \sim B$.
\end{itemize}

\vspace{\baselineskip} Savage's result was originally stated for qualitative probabilities on power sets, and the proof applies for qualitative probabilities on $\sigma$-algebras, but our model involves qualitative probabilities on algebras. We are therefore interested in the following generalization of Savage's theorem.\footnote{For alternate axioms for qualitative probabilities on algebras, see \cite{Chateauneuf1985} and \cite{Barbanel-Taylor1995}.}

\vspace{\baselineskip} \noindent \textsc{Theorem~WM (\citealp{Wakker1981}; \citealp{Marinacci1993}):} Fix a kitchen. If a qualitative probability $\succsim$ satisfies {\it fineness} and {\it tightness}, then there is a probability measure $\mu_0$ such that for each pair $A, B \in \mathcal{S}$, we have $A \succsim B$ if and only if $\mu_0(A) \geq \mu_0(B)$. In this case, $\mu_0$ is the unique such kitchen measure, and the range of $\mu_0$ is a dense subset of the unit interval.

\vspace{\baselineskip} We seek the stronger conclusion that preferences are represented by a kitchen measure, and accordingly introduce two axioms that correspond to the two conditions for kitchen measures.

\vspace{\baselineskip} \noindent \textsc{Definition:} Fix a kitchen. We say that a qualitative probability $\succsim$ satisfies
\begin{itemize}
\item {\it sliver nullity} if for each $S \in \mathcal{S}$, we have $S \sim \emptyset$ if and only if $S \in \mathcal{S}^\circ$; and

\item {\it knife continuity} if for each $\kappa \in \mathbb{K}$ and each $A \in \mathcal{S}$, both $\{ t \in [0, 1] | A \succsim \kappa_t \}$ and $\{ t \in [0, 1] | \kappa_t \succsim A\}$ are closed.
\end{itemize}

\vspace{\baselineskip} The first axiom requires no explanation. The second axiom requires that each knife is a continuous function from the unit interval to the collection of servings when the latter is associated with the coarsest topology for which upper contour sets and lower contour sets---that is, sets of the form $\{B \in \mathcal{S} | B \succsim A\}$ and $\{B \in \mathcal{S} | A \succsim B\}$, respectively---are closed. In other words, each knife is a path from the empty serving to the entire cake, and moreover one that is monotonic with respect to set inclusion.

It follows from \hyperlink{TheoremWM}{Theorem~WM} that Savage's axioms and our new axioms are together sufficient for kitchen measure representation, and it is not too difficult to show that they are also necessary. Less obvious, however, is that tightness is implied by the other axioms.

\hypertarget{Theorem2}{}
\vspace{\baselineskip} \noindent \textsc{Theorem 2:} Fix a kitchen. A qualitative probability $\succsim$ satisfies {\it fineness}, {\it sliver nullity}, and {\it knife continuity} if and only if there is a kitchen measure $\mu_0$ such that for each pair $A, B \in \mathcal{S}$, we have $A \succsim B$ if and only if $\mu_0(A) \geq \mu_0(B)$. In this case, $\mu_0$ is the unique such kitchen measure.

\vspace{\baselineskip} The proof is in \hyperlink{Appendix4}{Appendix~4}.

\hypertarget{Section5.2}{}
\subsection{The mediator and the adversary}

We conclude by investigating some strategic considerations raised by our formalization of protocols as pure strategies.

We begin by discussing a game for which the strategic analysis is not so interesting. Recall that in the game from \hyperlink{Section3.4}{Section~3.4}, an indifferent nature makes an unobserved choice of measure profile and thereafter the mediator is the only player. If we replace nature with an {\it adversary} whose ranking of outcomes is the opposite of the mediator's, then we can directly apply our analysis of the optimal protocol with a different interpretation: we analyze the strategy that is optimal for the mediator when the adversary selects his strategy after observing the mediator's. Unfortunately, repeating this exercise for the adversary yields little guidance: all strategy profiles are equivalent, because after observing the adversary's choice of measure profile, the mediator will immediately produce an $e$-proportional allocation. Altogether, we have an example of a two-player zero-sum game with neither a (pure strategy) value \citep{vonNeumann1928} nor a (pure strategy) Nash equilibrium \citep{Nash1951}. This is a familiar phenomenon; a classic example is Matching Pennies.

There is, however, a natural variant of the preceding game for which the strategic analysis is more interesting. In particular, consider the perfect information game where iteratively the mediator poses queries and the adversary responds.

\vspace{\baselineskip} \noindent \textsc{Definition:} Fix a setting. For each $e \in E$, the {\it adversary game given $e$} is the two player game defined as follows.
\begin{itemize}
\item There are two players: the {\it mediator} and the {\it adversary}. There are also histories partially ordered by precedence, as well as active players and actions associated with non-terminal histories and payoffs. The precedence order is such that each history and action at that history together determine an {\it immediate successor}, or an earliest history that follows. We define these remaining ingredients below, defining precedence entirely using immediate successors. The game has perfect information: all information sets are singletons.

\item A history is {\it odd} if it is non-terminal with an even number of predecessors, and {\it even} if it is non-terminal with an odd number of predecessors. The initial history is odd. At each odd history, the active player is the mediator and the set of available actions is $\mathcal{Q} \cup \mathcal{X}$. If the mediator selects a query, then the immediate successor is non-terminal. If the mediator selects an allocation, then the immediate successor is terminal.

\item At each even history, the active player is the adversary. Necessarily, the previous action was some query $\mathsf{q}$, and the adversary must select a response: the set of available actions is $\{ \mathsf{r}(\mathsf{q}|\mu) | \mu \in \mathbb{D} \}$. Regardless of the adversary's action, the next history is non-terminal.

\item Each odd history $h$ is identified with the finite sequence of actions that leads to $h$ from the initial history: there is a {\it (query) count} $c \in \mathbb{N}_0$ such that $h = (\mathsf{q}_1, \mathsf{r}_1, \mathsf{q}_2, \mathsf{r}_2, ..., \mathsf{q}_c, \mathsf{r}_c)$. In this case, the {\it $h$-domain} is $\mathbb{D}(h) \equiv \cap_{t \in \{1, 2, ..., c\}} \mathbb{D}(\mathsf{r}_t)$, where by convention $c=0$ implies $\mathbb{D}(h) = \mathbb{D}$.

\item A play is a maximal chain of histories. If a play ends at a terminal history after odd history $h$ with query count $c$, and if the allocation $X$ selected by the mediator has the property that for each $\mu \in \mathbb{D}(h)$ we have that $X$ is $(e|\mu)$-proportional, then the mediator's {\it payoff} is $-c$; otherwise it is $-\infty$. Observe that at an odd history $h$ with query count $c$, if $\mathbb{D}(h) = \emptyset$ because the adversary's responses are inconsistent, then the mediator can receive payoff $-c$ by selecting any allocation. The adversary's {\it payoff} is the negative of the mediator's payoff.
\end{itemize}
For each player, a {\it (pure) strategy} is a function that associates each history at which he is the active player with an action available at that history. We let $\Sigma_m$ denote the set of strategies for the mediator and let $\Sigma_a$ denote the set of strategies for the adversary. Finally, for each strategy profile $(\sigma_m, \sigma_a) \in \Sigma_m \times \Sigma_a$, we let $\mathbb{P}(\sigma_m, \sigma_a)$ denote the associated payoff to the mediator, which the mediator seeks to maximize and the adversary seeks to minimize.\footnote{We use {\it payoffs} instead of {\it utilities} to emphasize that we are not considering randomization: the payoffs are not derived from preferences over lotteries and we are not considering mixed strategies.}

\vspace{\baselineskip} We are interested in analyzing the adversary games using the classic notions of value \citep{vonNeumann1928} and equilibrium \citep{Nash1951}.

\vspace{\baselineskip} \noindent \textsc{Definition:} Fix a setting, let $e \in E$, and consider the associated adversary game.
\begin{itemize}
\item If $\sup_{\sigma_m \in \Sigma_m} \inf_{\sigma_a \in \Sigma_a} \mathbb{P}(\sigma_m, \sigma_a) = \inf_{\sigma_a \in \Sigma_a} \sup_{\sigma_m \in \Sigma_m} \mathbb{P}(\sigma_m, \sigma_a)$, then this is {\it the game's value}; otherwise {\it the game has no value}.

\item A strategy profile $(\sigma_m, \sigma_a)$ is a {\it (pure strategy) Nash equilibrium} if (i)~for each $\sigma'_m \in \Sigma_m$, $\mathbb{P}(\sigma_m, \sigma_a) \geq \mathbb{P}(\sigma'_m, \sigma_a)$, and (ii)~for each $\sigma'_a \in \Sigma_a$, $\mathbb{P}(\sigma_m, \sigma_a) \leq \mathbb{P}(\sigma_m, \sigma'_a)$.
\end{itemize}

\vspace{\baselineskip} In each adversary game, each player has an infinite set of strategies without any further mathematical structure that the payoff function is required to respect. While it is possible to guarantee that such a game has a value if the payoff function is {\it concave-convexlike} (\citealp{Fan1953}; \citealp{Sion1958}), that is not the case here, and we are aware of no general theorem that guarantees the adversary games have values. Fortunately, we can do so using our earlier analysis.

\hypertarget{Theorem3}{}
\vspace{\baselineskip} \noindent \textsc{Theorem 3:} Fix a setting and let $e \in E$. If all entitlements are rational, then the game has a value and this value is achieved in a Nash equilibrium. If there is an irrational entitlement, then the game's value is $-\infty$ but the game has no Nash equilibrium.

\vspace{\baselineskip} The proof is in \hyperlink{Appendix5}{Appendix~5}. The first conclusion follows directly from the existence of a bounded protocol when all entitlements are rational \citep{Steinhaus1948}, but the second conclusion does not follow directly from our result that there is no bounded protocol when some entitlement is irrational (\hyperlink{Corollary1}{Corollary~1}): for each cost we must construct an adversary strategy that guarantees that cost, and for this we require our deeper insights about deficient partition records and ultraresponses (\hyperlink{Lemma4}{Lemma~4}).

\hyperlink{Theorem3}{Theorem~3} reveals that so far as value and equilibrium are concerned, any adversary game with an irrational entitlement is effectively a one-player game: the mediator is effectively a dummy player and the adversary can effectively select any cost $c \in \mathbb{N}$ to transfer to himself from the mediator. In this way, we can articulate the idea that this is a hard problem for the mediator using classic notions from game theory without appealing to randomization.

\hypertarget{Section6}{}
\section{Discussion}

We have investigated the communication costs of privatizing some jointly-owned property that is non-homogeneous and divisible in accordance with well-defined ownership shares, and found that even under the first-best assumption that all agents are honest, the problem is in general impractical: there may be no deadline by which a solution can be guaranteed. Moreover, we have shown that communication costs are caused at least in part by the problem's clonage, or the least common denominator of the entitlements.

When it is impractical to achieve an objective exactly, it is natural to approximate, and this direction has been investigated in the context of cake division protocols for several fairness notions including no-envy \citep{Branzei-Nisan2022}. To pursue approximate $e$-proportionality when one of the entitlements is irrational, a simple idea is to approximate the entitlements with rational numbers, and this idea invites the use of continued fractions because in a suitable sense they provide the best rational approximations of irrational numbers \citep{Hardy-Wright2008}.\footnote{We thank Fedor Sandomirskiy for this suggestion.} Indeed, though it would be impractical to exactly respect the entitlement profile $(\frac{1}{\pi}, 1-\frac{1}{\pi})$, perhaps in this case it would be fine to respect $(\frac{1}{3}, \frac{2}{3})$, $(\frac{7}{22}, \frac{15}{22})$, or $(\frac{106}{333}, \frac{227}{333})$. We leave a more thorough investigation of this problem for future work.

\appendix
\hypertarget{Appendix1}{}
\setcounter{secnumdepth}{0}
\section{Appendix 1}

In this appendix, we prove \hyperlink{Lemma3}{Lemma~3}. To do so, we first prove \hyperlink{Lemma2}{Lemma~2}, which provides a convenient description of when a record is an ultraresponse.

\hypertarget{Lemma2}{}
\vspace{\baselineskip} \noindent \textsc{Lemma 2:} Fix a setting. For each valid partition record $(\mathcal{P}, \alpha)$, each query $\mathsf{q} = (i, \kappa, S, p)$, and each record $(\mathcal{P}^+, \alpha^+)$, $(\mathcal{P}^+, \alpha^+) \in \mathcal{R}(\mathsf{q}|\mathcal{P}, \alpha)$ if and only if it satisfies the following conditions.
\begin{itemize}
\item {\it Cut compatibility.} There is $\mu \in \mathbb{D}(\mathcal{P}, \alpha)$ such that (i)~$\mathcal{P}^+$ is the collection of atoms in the subalgebra of $\mathcal{S}$ generated by $\mathcal{P} \cup \{S, S_{\mathsf{q}|\mu_i}\}$, and (ii)~$\alpha^{+\uparrow}_i(S_{\mathsf{q}|\mu_i}) = p \cdot \alpha^{+\uparrow}_i(S)$.

\item {\it Appraisal compatibility.} For each $j \in N$ and each $A \in \mathcal{P}$, $\sum_{\{B \in \mathcal{P}^+ | B \subseteq A\}} \alpha^+_j(B) = \alpha_j(A)$.

\item {\it Sliver compatibility.} For each $j \in N$ and each $A \in \mathcal{P}^+$, $\alpha^+_j(A) = 0$ if and only if $A \in \mathcal{S}^\circ$.
\end{itemize}

\vspace{\baselineskip} \noindent \textsc{Proof:} It is straightforward to verify that the conditions are necessary using \hyperlink{Lemma1}{Lemma~1}; we omit the argument. To see that the conditions are sufficient, assume the hypotheses and let $\mu$ be the member of $\mathbb{D}(\mathcal{P}, \alpha)$ promised by cut compatibility.

First, we construct $\mu^* \in \mathbb{D}$. Indeed, fix $j \in N$. We define $\mu^*_j:\mathcal{S} \to [0,1]$ to be such that for each $A \in \mathcal{S}$, $\mu^*_j(A) \equiv \sum_{B \in \mathcal{P}^+ \backslash \mathcal{S}^\circ} \alpha^+_j(B) \cdot \frac{\mu_j(A \cap B)}{\mu_j(B)}$, which for emphasis does not involve division by zero because $B \in \mathcal{P}^+ \backslash \mathcal{S}^\circ$ implies $\mu_j(B) > 0$. By null slivers, appraisal compatibility, and the fact that $(\mathcal{P}, \alpha)$ is valid, we have $\sum_{B \in \mathcal{P}^+ \backslash \mathcal{S}^\circ} \alpha^+_j(B) = \sum_{B \in \mathcal{P}^+} \alpha^+_j(B) = \sum_{A \in \mathcal{P}} \alpha_j(A) = 1$, so $\mu^*_j$ is a convex combination of probability measures that satisfy knife divisibility, from which it is straightforward to show that $\mu^*_j$ is a probability measure that satisfies knife divisibility. Moreover, since $\mu_j$ satisfies null slivers and $\alpha^+_j$ satisfies sliver compatibility, thus $\mu^*_j$ satisfies null slivers. Altogether, then, $\mu^*_j$ is a kitchen measure. Since $j \in N$ was arbitrary, thus $\mu^* \in \mathbb{D}$, as desired.

To conclude, we prove that $(\mathcal P^+,\alpha^+) = \mathsf{r}(\mathsf{q}|\mu^*,\mathcal P, \alpha)$. To begin, by definition of $\mu^*_i$, the fact that each member of $\mathcal{P}^+$ is either contained in $S_{\mathsf{q}|\mu_i}$ or $C \backslash S_{\mathsf{q}|\mu_i}$, sliver compatibility, and the definition of subalgebra extension, respectively, we have
\begin{align*}
\mu_i^*(S_{\mathsf{q}|\mu_i}) &= \sum_{B \in \mathcal{P}^+ \backslash \mathcal{S}^\circ} \alpha^+_i(B) \cdot \frac{\mu_i(S_{\mathsf{q}|\mu_i} \cap B)}{\mu_i(B)}
\\ &= \sum_{\{B \in \mathcal{P}^+ \backslash \mathcal{S}^\circ | B \subseteq S_{\mathsf{q}|\mu_i}\}} \alpha^+_i(B)
\\ &= \sum_{\{B \in \mathcal{P}^+ | B \subseteq S_{\mathsf{q}|\mu_i}\}} \alpha^+_i(B)
\\ &= \alpha^{+\uparrow}_i(S_{\mathsf{q}|\mu_i}).
\end{align*}
By the same argument we have $\mu_i^*(S \backslash S_{\mathsf{q}|\mu_i}) = \alpha_i^{+\uparrow}(S \backslash S_{\mathsf{q}|\mu_i})$, so by the additivity of both $\mu^*_i$ and $\alpha_i^{+\uparrow}$ we have $\mu_i^*(S) = \alpha_i^{+\uparrow}(S)$. Thus by cut compatibility, we have
\begin{align*}
\mu^*_i(S_{\mathsf{q}|\mu_i}) &= \alpha_i^{+\uparrow}(S_{\mathsf{q}|\mu_i})
\\ &= p \cdot \alpha_i^{+\uparrow}(S)
\\ &= p \cdot \mu^*_i(S).
\end{align*}
Moreover, fix $t \in [0,\tau(\mathsf{q}|\mu_i))$. First, by definition of $\tau(\mathsf{q}|\mu_i)$ we have $S \cap \kappa_t \subseteq S_{\mathsf{q}|\mu_i}$, so for each $B \in \mathcal{P}^+$ we have $\mu^*_i((S \cap \kappa_t) \cap B) \leq \mu^*_i(S_{\mathsf{q}|\mu_i} \cap B)$. Second, by definition of $\tau(\mathsf{q}|\mu_i)$ and \hyperlink{Lemma1}{Lemma~1} we have $\mu_i(S \cap \kappa_t) < p \cdot \mu_i(S) = \mu_i(S_{\mathsf{q}|\mu_i})$, so there is $B \in \mathcal{P}^+$ such that $\mu_i((S \cap \kappa_t) \cap B) < \mu_i(S_{\mathsf{q}|\mu_i} \cap B)$ and thus $\mu^*_i((S \cap \kappa_t) \cap B) < \mu^*_i(S_{\mathsf{q}|\mu_i} \cap B)$. Thus by additivity of $\mu^*_i$, we have $\mu_i^*(S \cap \kappa_t) < \mu^*_i(S_{\mathsf{q}|\mu_i}) = p \cdot \mu_i^*(S)$. Since $t \in [0,\tau(\mathsf{q}|\mu_i))$ was arbitrary, altogether we have $S_{\mathsf{q}|\mu_i^*} = S_{\mathsf{q}|\mu_i}$, from which it is straightforward to verify that $(\mathcal P^+,\alpha^+) = \mathsf{r}(\mathsf{q}|\mu^*,\mathcal P, \alpha)$.~$\blacksquare$

\vspace{\baselineskip} We conclude this appendix by providing \hyperlink{Lemma3}{Lemma~3}.

\hypertarget{Lemma3}{}
\vspace{\baselineskip} \noindent \textsc{Lemma 3 (Restated):} Fix a setting with two agents in $N = \{1, 2\}$ and let $e \in E$. For each level~$\ell$, each valid partition record $(\mathcal{P}, \alpha)$ that is $\ell$-deficient, and each $\textsf{q} \in \mathcal{Q}$ such that the serving argument of $\mathsf{q}$ belongs to $\mathcal{P}$, there is $(\mathcal{P}^+, \alpha^+) \in \mathcal{R}(\mathsf{q}|\mathcal{P}, \alpha)$ that is $\lfloor \sqrt{\ell/2} \rfloor$-deficient.

\vspace{\baselineskip} \noindent \textsc{Proof:} Let $\ell$, $(\mathcal{P}, \alpha)$, and $\mathsf{q}$ satisfy the hypotheses. Several concepts in this proof come in pairs: we use $1$ and $2$ to distinguish the agents, $'$ and $''$ to distinguish the two servings in $\mathcal{P}^+ \backslash \mathcal{P}$, and $^+$ to distinguish objects for the new partition from associated objects for the starting partition. To complete the proof, we select an arbitrary kitchen measure profile $\mu$ from a nonempty set, then construct the desired $(\mathcal{P}^+, \alpha^+)$.

\vspace{\baselineskip} \noindent \textsc{Step 1:} {\it Introduce concepts and notation to be used for the remaining steps.}

\vspace{\baselineskip} \noindent \textsc{Step 1a:} {\it Introduce preliminary notation, then construct (i)~the servings $B'$ and $B''$, (ii)~the partition $\mathcal{P}^+$, and (iii)~the appraisal $\alpha^+_1$. Moreover, proceed under the assumption that $B'$ and $B''$ are not slivers.}

\vspace{\baselineskip} First, we introduce preliminary notation. Define $\ell^+ \equiv \lfloor \sqrt{\ell/2} \rfloor$, let $s \equiv |\mathcal{P}|$ denote the number of slices in $\mathcal{P}$, and index the members of $\mathcal{P}$ by $\{A_j\}_{j \in \{1, 2, ..., s-1\}} \cup \{B\}$ such that~$B$ is the serving argument of $\mathsf{q}$. Furthermore, denote all the arguments of $\mathsf{q}$ by $(1, \kappa, B, p_1)$, so that (re-indexing if necessary) $\mathsf{q}$ asks agent $1$ to cut, and let $2$ denote the other agent. Observe that such a re-indexing is without loss of generality even though our notion of hyperallocation does not treat agents symmetrically: for each level $\ell'$, a partition record is $\ell'$-deficient before re-indexing if and only if it is $\ell'$-deficient after re-indexing. Define $v_1 \equiv p_1 \cdot \alpha_1(B)$, so that $\mathsf{q}$ effectively asks $1$ to use~$\kappa$ to construct a subset of $B$ worth $v_1$.

Second, we construct $B'$, $B''$, $\mathcal{P}^+$, and $\alpha^+_1$. To begin, since $(\mathcal{P}, \alpha)$ is valid, thus we can select an arbitrary $\mu \in \mathbb{D}(\mathcal{P}, \alpha)$; we remark that for the ultraresponse $(\mathcal{P}^+, \alpha^+)$ that we will construct, both $\mathcal{P}^+$ and $\alpha^+_1$ will be consistent with $\mu_1$. We define $B' \equiv S_{\mathsf{q}|\mu_1}$ and define $B'' \equiv B \backslash B'$. If $B' \in \mathcal{S}^\circ$ or $B'' \in \mathcal{S}^\circ$, then for $(\mathcal{P}^+, \alpha^+) \equiv \mathsf{r}(\mathsf{q}|\mu, \mathcal{P}, \alpha)$ we have $\mathbb{D}(\mathcal{P}, \alpha) = \mathbb{D}(\mathcal{P}^+, \alpha^+)$, from which the desired conclusion directly follows; thus let us assume $B' \not \in \mathcal{S}^\circ$ and $B'' \not \in \mathcal{S}^\circ$. Define $\mathcal{P}^+ \equiv (\mathcal{P} \backslash \{B\}) \cup \{B', B''\}$, and observe that this is indeed a partition. Finally, define $\alpha^+_1 \equiv {\mu_1}\restriction_{\mathcal{P}^+}$. Observe that by construction, (i)~for each $A \in \mathcal{P}^+ \backslash \{B', B''\}$ we have $\alpha^+_1(A) = \alpha_1(A)$, (ii)~$\alpha^+_1(B') = v_1$, and (iii)~$\alpha^+_1(B'') = \alpha^+_1(B) - v_1$.

Finally, we highlight an abuse of notation: throughout this proof, we always use the same notation for an appraisal, its subalgebra extension, and its hyperallocation extension. For example, instead of distinguishing between $\alpha^+_1$, $\alpha^{+\uparrow}_1$, and $\alpha^{+\uparrow \uparrow}_1$, we simply write $\alpha^+_1$ for all three functions, with the particular function we are using determined not by notation but by context.

\vspace{\baselineskip} \noindent \textsc{Step 1b:} {\it Introduce the possible values that $2$ assigns to $B'$, their associated appraisals and records, and related concepts.}

\vspace{\baselineskip} First, we define the set of possible values that~$2$ assigns to $B'$ (given that neither $B'$ nor $B''$ is a sliver), its supremum, and its closure. In particular, (i)~define $v^{\sup} \equiv \alpha_2(B)$, (ii)~define $V \equiv (0, v^{\sup})$ to be the set of possible values that~$2$ assigns to $B'$, and (iii)~define $\overline{V} \equiv [0, v^{\sup}]$. Since $B'$ and $B''$ are not slivers, thus $B = B' \cup B''$ is not a sliver, so as $(\mathcal{P}, \alpha)$ is valid we have that $v^{\sup} = \alpha_2(B) > 0$.

Second, for each $v \in \overline{V}$ we define (i)~the $\mathcal{P}^+$-appraisal $\alpha_2^v: \mathcal{P}^+ \to [0, 1]$, and (ii)~the record $\mathsf{r}_v$, as follows:
\begin{itemize}
\item $\alpha_2^v$ is defined by (i)~for each $A \in \mathcal{P}^+ \backslash \{B', B''\}$, $\alpha^v_2(A) = \alpha_2(A)$, (ii)~$\alpha^v_2(B') = v$, and (iii)~$\alpha^v_2(B'') = \alpha_2(B) - v$, and

\item $\mathsf{r}_v \equiv (\mathcal{P}^+, (\alpha^+_1, \alpha^v_2))$.
\end{itemize}
Observe that by \hyperlink{Lemma2}{Lemma~2}, for each $v \in V$ we have $\mathsf{r}_v \in \mathcal{R}(\mathsf{q}|\mathcal{P}, \alpha)$. By contrast, for each $v \in \{0, v^{\sup}\}$, $\mathsf{r}_v$ is not valid because it assigns zero to a serving that is not a sliver, and thus $\mathsf{r}_v \not \in \mathcal{R}(\mathsf{q}|\mathcal{P}, \alpha)$. Even so, it is convenient to use the values in $\{0, v^{\sup}\}$ for some of our arguments.

\vspace{\baselineskip} \noindent \textsc{Step 1c:} {\it Introduce weights, lines, and candidates.}

\vspace{\baselineskip} First, for each hyperallocation $X \in \mathcal{X}_{\ell^+|\mathcal{P}^+}$, we write the value that $2$ assigns to $X$ as a function of the value that he assigns to $B'$. In particular, for each $X \in \mathcal{X}_{\ell^+|\mathcal{P}^+}$, we define the weights $w'(X), w''(X) \in \{0, 1, ..., r(X)\}$ and the line $\mathcal{L}_X: \overline{V} \to [0, 1]$, as follows:
\begin{align*}
w'(X) &\equiv |\{j \in \{1, 2, ..., r(X)\} | B' \subseteq X_j \}|,
\\ w''(X) &\equiv |\{j \in \{1, 2, ..., r(X)\} | B'' \subseteq X_j \}|, \text{ and}
\\ \mathcal{L}_X(v) &\equiv \alpha_2^v(X)
\\ &= \Big[\frac{[\sum_{j = 1}^{r(X)}\alpha_2(X_j \backslash B)] + w''(X) \cdot v^{\sup}}{r(X)}\Big] + \Big[\frac{w'(X) - w''(X)}{r(X)} \Big] \cdot v.
\end{align*}
We say that $X$ is (i)~{\it balanced} if $w'(X) = w''(X)$, (ii)~{\it $B'$-heavy} if $w'(X) > w''(X)$, and (iii)~{\it $B''$-heavy} if $w'(X) < w''(X)$. Observe that a balanced $X$ has a flat $X$-line, a $B'$-heavy $X$ has an increasing $X$-line, and a $B''$-heavy $X$ has a decreasing $X$-line.

Second, for each $X \in \mathcal{X}_{\ell^+|\mathcal{P}^+}$, we say that $X$ is a {\it candidate} if $\alpha^+_1(X) \geq e_1$. To complete the proof, we construct $v^* \in V$ such that for each candidate $X$ we have $\mathcal{L}_X(v^*) = \alpha_2^{v^*}(X) > e_1$, as this directly implies that $\mathsf{r}_{v^*}$ is $\ell^+$-deficient. Intuitively, if $2$ assigns such a value $v^*$ to $B'$, then for each hyperallocation that is a candidate for being satisfactory in the sense that $1$ measures his own hyperserving to be worth at least his own entitlement, the hyperallocation fails to be satisfactory because $2$ measures $1$'s hyperserving to be worth more than $1$'s entitlement---and thus measures his own hyperserving to be worth less than his own entitlement.

\vspace{\baselineskip} \noindent \textsc{Step 2:} {\it For each balanced candidate $X^+ \in \mathcal{X}_{\ell^+|\mathcal{P}^+}$ and each $v \in \overline{V}$, $\mathcal{L}_{X^+}(v) > e_1$.}

\vspace{\baselineskip} Let $X^+$ and $v$ satisfy the hypotheses. For convenience, let us write $r^+ \equiv r(X^+)$, $\alpha^+_2 \equiv \alpha^v_2$, $w' \equiv w'(X^+)$, and $w'' \equiv w''(X^+)$. Since $X^+$ is balanced, thus $w' = w''$. Observe that $r^+ \leq \ell^+ = \lfloor \sqrt{\ell/2} \rfloor \leq \ell$.

Let $X \in \mathcal{X}_{\ell|\mathcal{P}}$ with $r(X) = r^+$ be such that for each $j \in \{1, 2, ..., r^+\}$, (i)~$j \leq w'$ implies $X_j \equiv (X^+_j \backslash B) \cup B$, and (ii)~$j > w'$ implies $X_j \equiv X^+_j \backslash B$. Then for each $i \in N$, we have
\begin{align*}
\alpha_i(X) &= \Big[ \sum_{j = 1}^{r^+} \frac{\alpha_i(X^+_j \backslash B)}{r^+} \Big] + w' \cdot \frac{\alpha_i(B)}{r^+}
\\ &= \Big[ \sum_{j = 1}^{r^+} \frac{\alpha^+_i(X^+_j \backslash B)}{r^+} \Big] + w' \cdot \frac{\alpha^+_i(B')}{r^+} + w' \cdot \frac{\alpha^+_i(B'')}{r^+}
\\ &= \Big[ \sum_{j = 1}^{r^+} \frac{\alpha^+_i(X^+_j \backslash B)}{r^+} \Big] + w' \cdot \frac{\alpha^+_i(B')}{r^+} + w'' \cdot \frac{\alpha^+_i(B'')}{r^+}
\\ &= \alpha^+_i(X^+).
\end{align*}
Since $X^+$ is a candidate, thus we have $\alpha_1(X) = \alpha^+_1(X^+) \geq e_1$, so since $(\mathcal{P}, \alpha)$ is $\ell$-deficient we have $\mathcal{L}_{X^+}(v) = \alpha^+_2(X^+) = \alpha_2(X) > e_1$, as desired.

\vspace{\baselineskip} \noindent \textsc{Step 3:} {\it For each $B'$-heavy candidate $X^+ \in \mathcal{X}_{\ell^+|\mathcal{P}^+}$, $\mathcal{L}_{X^+}(v^{\sup}) > e_1$.}

\vspace{\baselineskip} Let $X^+$ satisfy the hypotheses. For convenience, let us write  $r^+ \equiv r(X^+)$, $\alpha^+_2 \equiv \alpha^{v^{\sup}}_2$, $w' \equiv w'(X^+)$, and $w'' \equiv w''(X^+)$. Since $X^+$ is $B'$-heavy, thus $w' > w''$. Observe that $r^+ \leq \ell^+ = \lfloor \sqrt{\ell/2} \rfloor \leq \ell$.

Let $X \in \mathcal{X}_{\ell|\mathcal{P}}$ with $r(X) = r^+$ be such that for each $j \in \{1, 2, ..., r^+\}$, (i)~$j \leq w'$ implies $X_j \equiv (X^+_j \backslash B) \cup B$, and (ii)~$j > w'$ implies $X_j \equiv X^+_j \backslash B$. Then for each $i \in N$, we have
\begin{align*}
\alpha_i(X) &= \Big[ \sum_{j = 1}^{r^+} \frac{\alpha_i(X^+_j \backslash B)}{r^+} \Big] + w' \cdot \frac{\alpha_i(B)}{r^+}
\\ &= \Big[ \sum_{j = 1}^{r^+} \frac{\alpha^+_i(X^+_j \backslash B)}{r^+} \Big] + w' \cdot \frac{\alpha^+_i(B')}{r^+} + w' \cdot \frac{\alpha^+_i(B'')}{r^+}
\\ &\geq \Big[ \sum_{j = 1}^{r^+} \frac{\alpha^+_i(X^+_j \backslash B)}{r^+} \Big] + w' \cdot \frac{\alpha^+_i(B')}{r^+} + w'' \cdot \frac{\alpha^+_i(B'')}{r^+}
\\ &= \alpha^+_i(X^+).
\end{align*}
Moreover, since $\alpha^+_2(B'') = 0$, thus the inequality above holds with equality for agent $i = 2$. Since $X^+$ is a candidate, thus we have $\alpha_1(X) \geq \alpha^+_1(X^+) \geq e_1$, so since $(\mathcal{P}, \alpha)$ is $\ell$-deficient we have $\mathcal{L}_{X^+}(v^{\sup}) = \alpha^+_2(X^+) = \alpha_2(X) > e_1$, as desired.

\vspace{\baselineskip} \noindent \textsc{Step 4:} {\it For each $B''$-heavy candidate $X^+ \in \mathcal{X}_{\ell^+|\mathcal{P}^+}$, $\mathcal{L}_{X^+}(0) > e_1$.}

\vspace{\baselineskip} The argument is the same as that for Step~3, except we use $0$ instead of $v^{\sup}$ and we transpose the roles of $w'$ and $w''$.

\vspace{\baselineskip} \noindent \textsc{Step 5:} {\it Construct $v', v'' \in \overline{V}$ such that
\begin{align*}
v' &= \min \{ v \in \overline{V} | \text{ for each $B'$-heavy candidate $X$, $\mathcal{L}_X(v) \geq e_1$} \}, \text{ and}
\\ v'' &= \max \{ v \in \overline{V} | \text{ for each $B''$-heavy candidate $X$, $\mathcal{L}_X(v) \geq e_1$} \},
\end{align*}
and such that moreover $v' < v^{\sup}$ and $v'' > 0$.}

\vspace{\baselineskip} Before we begin, we observe that if we allow all values in $\overline{V}$, then $v'$ is the tightest value threshold above which no $B'$-heavy candidate is accepted and $v''$ is the tightest value threshold below which no $B''$-heavy candidate is accepted; thus these are the values mentioned in the final paragraph of the proof sketch.

First, we construct $v'$. If there is no $B'$-heavy candidate, then define $v' \equiv 0$ and we are done; thus let us assume there is a $B'$-heavy candidate. In this case, by Step~3, for each $B'$-heavy candidate $X$, $\mathcal{L}_X$ is a strictly increasing line with $\mathcal{L}_X(v^{\sup}) > e_1$, so $v(X) \equiv \min \{v \in \overline{V}| \mathcal{L}_X(v) \geq e_1\}$ is well-defined and less than $v^{\sup}$; thus as there are finitely many $B'$-heavy candidates, $v' \equiv \max \{v(X) | \text{ $X$ is a $B'$-heavy candidate} \}$ is well-defined and satisfies the desired properties.

Second, we construct $v''$. The construction is analogous to the construction of $v'$, using $B''$-heavy candidates and Step~4 instead of $B'$-heavy candidates and Step~3; we omit the details.

\vspace{\baselineskip} \noindent \textsc{Step 6:} {\it Proceed under the assumption that there are a $B'$-heavy candidate and a $B''$-heavy candidate.}

\vspace{\baselineskip} If there is no $B'$-heavy candidate and there is no $B''$-heavy candidate, then define $v^* = \frac{v^{\sup}}{2}$; by Step~2, for each candidate $X$ we have $\mathcal{L}_X(v^*) > e_1$, so we are done. If there is a $B'$-heavy candidate but there is no $B''$-heavy candidate, then define $v^* = \frac{v'+v^{\sup}}{2}$; by Step~2 and Step~5, for each candidate $X$ we have $\mathcal{L}_X(v^*) > e_1$, so we are done. If there is a $B''$-heavy candidate but there is no $B'$-heavy candidate, then define $v^* = \frac{v''}{2}$; by Step~2 and Step~5, for each candidate $X$ we have $\mathcal{L}_X(v^*) > e_1$, so we are done. Thus let us assume there are a $B'$-heavy candidate and a $B''$-heavy candidate.

\vspace{\baselineskip} \noindent \textsc{Step 7:} {\it We have $v' < v''$.} 

\vspace{\baselineskip} Assume for contradiction that $v' \geq v''$ and let $v^* \in [v'', v']$. Then by Step~5, we have $v^{\sup} > v' \geq v'' > 0$, so $v' \in V$ and $v'' \in V$. Moreover, since there are finitely many candidates, thus by definition of $v'$ and $v''$, there are $B'$-heavy candidate $X' \in \mathcal{X}_{\ell^+|\mathcal{P}^+}$ and $B''$-heavy candidate $X'' \in \mathcal{X}_{\ell^+|\mathcal{P}^+}$ such that $\mathcal{L}_{X'}(v^*) \leq \mathcal{L}_{X'}(v') = e_1$ and $\mathcal{L}_{X''}(v^*) \leq \mathcal{L}_{X''}(v'') = e_1$. For convenience, let us write $r' \equiv r(X')$, $r'' \equiv r(X'')$, and $\alpha^+_2 \equiv \alpha^{v^*}_2$.

Intuitively, we construct a hyperallocation that (i)~consists of some copies of $X'$ and some copies of $X''$, (ii)~involves a grand total of at most $\ell$ replicas, and (iii)~includes the same number of $B'$ slices and $B''$ slices. This $(\ell|\mathcal{P}^+)$-hyperallocation is equivalent to an $(\ell|\mathcal{P})$-hyperallocation that we refer to as $X$.

We begin by using $X'$ and $X''$ to define the weight lists $(w'_S)_{S \in \mathcal{P}^+}$ and $(w''_S)_{S \in \mathcal{P}^+}$ and the weight differences $\delta'$ and $\delta''$, then using these notions to define the number of replicas~$r$ and the weight list $(w_S)_{S \in \mathcal{P}^+}$. In particular:
\begin{itemize}
\item for each $S \in \mathcal{P}^+$, $w'_S \equiv |\{j \in \{1, 2, ..., r'\} | S \subseteq X'_j\}|$,

\item for each $S \in \mathcal{P}^+$,  $w''_S \equiv |\{j \in \{1, 2, ..., r''\} | S \subseteq X''_j\}|$,

\item $\delta' \equiv [w'_{B'} - w'_{B''}]$,

\item $\delta'' \equiv [w''_{B''} - w''_{B'}]$,

\item $r \equiv \delta'' r' + \delta' r''$, and

\item for each $S \in \mathcal{P}^+$, $w_S \equiv \delta'' w'_S + \delta' w''_S$.
\end{itemize}
We use these objects to establish our contradiction. Intuitively, we construct a new hyperallocation that consists of $\delta''$ copies of $X'$ and $\delta'$ copies of $X''$.\footnote{We remark that at this point, an alternative path leads to a slightly tighter but less simple bound. In particular, if we let $g$ denote the greatest common divisor of $\delta'$ and $\delta''$, then we could complete an analogous argument by making only $\frac{\delta''}{g}$ copies of $X'$ and $\frac{\delta'}{g}$ copies of $X''$. We choose to pursue the slightly looser but simpler bound.}

Next, we make five observations. First, since $r', r'' \in \{0, 1, ..., \ell^+\}$, thus for each $S \in \mathcal{P}^+$ we have $w'_S, w''_S \in \{0, 1, ..., \ell^+\}$. Second, we have $\sum_{S \in \mathcal{P}^+} w'_S \cdot \frac{\alpha^+_2(S)}{r'} = \mathcal{L}_{X'}(v^*) \leq e_1$ and $\sum_{S \in \mathcal{P}^+} w''_S \cdot \frac{\alpha^+_2(S)}{r''} = \mathcal{L}_{X''}(v^*) \leq e_1$. Third, we have $\delta', \delta'', r', r'' \in \{0, 1, ..., \ell^+\}$ and thus $r \in \{0, 1, ..., 2(\ell^+)^2\}$, and moreover $\ell^+ = \lfloor \sqrt{\ell/2} \rfloor$, so $r \in \{0, 1, ..., \ell\}$. Fourth, for each $S \in \mathcal{P}^+$, since $w'_S \in \{0, 1, ..., r'\}$ and $w''_S \in \{0, 1, ..., r''\}$, thus $w_S \in \{0, 1, ..., r\}$. Fifth, we have
\begin{align*}
w_{B'} &= [w''_{B''} - w''_{B'}] w'_{B'} + [w'_{B'} - w'_{B''}] w''_{B'}
\\ &= w''_{B''} w'_{B'} - w''_{B'} w'_{B''}
\\ &= [w''_{B''} - w''_{B'}] w'_{B''} + [w'_{B'} - w'_{B''}] w''_{B''}
\\ &= w_{B''}.
\end{align*}
Define $w_B \equiv w_{B'} = w_{B''}$.

Finally, for each $j \in \{1, 2, ..., r\}$, define $X_j \equiv \cup \{ S \in \mathcal{P} | j \leq w_S\}$. Since $r \leq \ell$, thus $X \in \mathcal{X}_{\ell|\mathcal{P}}$. Using the above definitions and observations, for each $i \in N$ we have
\begin{align*}
\alpha_i(X) &= \sum_{j=1}^r \frac{\alpha_i(X_j)}{r}
\\ &= \frac{\sum_{A \in \mathcal{P} \backslash \{B\}} w_A \alpha_i(A)}{r} + \frac{w_B \alpha_i(B)}{r}
\\ &= \frac{\sum_{A \in \mathcal{P}^+ \backslash \{B', B''\}} w_A \alpha^+_i(A)}{r} + \frac{w_B \alpha_i^+(B)}{r}
\\ &= \frac{\sum_{A \in \mathcal{P}^+ \backslash \{B', B''\}} w_A \alpha^+_i(A)}{r} + \frac{w_B \alpha^+_i(B') + w_B \alpha^+_i(B'')}{r}
\\ &= \frac{\sum_{A \in \mathcal{P}^+ \backslash \{B', B''\}} w_A \alpha^+_i(A)}{r} + \frac{w_{B'} \alpha^+_i(B') + w_{B''} \alpha^+_i(B'')}{r}
\\ &= \frac{\sum_{A \in \mathcal{P}^+} w_A \alpha^+_i(A)}{r}
\\ &= \frac{\sum_{A \in \mathcal{P}^+} (\delta'' w'_A + \delta' w''_A) \alpha^+_i(A)}{r}
\\ &= \delta'' \frac{\sum_{A \in \mathcal{P}^+} w'_A \alpha^+_i(A)}{r} + \delta' \frac{\sum_{A \in \mathcal{P}^+} w''_A \alpha^+_i(A)}{r}
\\ &= (\frac{\delta''r'}{r}) \frac{\sum_{A \in \mathcal{P}^+} w'_A \alpha^+_i(A)}{r'} + (\frac{\delta'r''}{r}) \frac{\sum_{A \in \mathcal{P}^+} w''_A \alpha^+_i(A)}{r''}
\\ &= (\frac{\delta''r'}{\delta'' r' + \delta' r''}) \alpha^+_i(X') + (\frac{\delta'r''}{\delta'' r' + \delta' r''}) \alpha^+_i(X'').
\end{align*}
But then (i)~since $X'$ and $X''$ are candidates we have $\alpha_1(X) \geq e_1$, and (ii)~since $\alpha^+_2(X') = \mathcal{L}_{X'}(v^*) \leq e_1$ and $\alpha^+_2(X'') = \mathcal{L}_{X''}(v^*) \leq e_1$ we have $\alpha_2(X) \leq e_1$, contradicting that $(\mathcal{P}, \alpha)$ is $\ell$-deficient.

\vspace{\baselineskip} \noindent \textsc{Step 8:} {\it Conclude.}

\vspace{\baselineskip} By Step~7, we have $v' < v''$. Define $v^* \equiv \frac{v' + v''}{2}$. First, by Step~2, for each balanced candidate $X$ we have $\mathcal{L}_X(v^*) > e_1$. Second, using the definition of $v'$, it is easy to verify that for each $B'$-heavy candidate $X$ we have $\mathcal{L}_X(v^*) > e_1$. Finally, using the definition of $v''$, it is easy to verify that for each $B''$-heavy candidate $X$ we have $\mathcal{L}_X(v^*) > e_1$. Altogether, then, for each candidate $X$ we have $\mathcal{L}_X(v^*) > e_1$, so $\mathsf{r}_{v^*}$ is $\ell^+$-deficient, as desired.~$\blacksquare$

\hypertarget{Appendix2}{}
\setcounter{secnumdepth}{0}
\section{Appendix 2}

In this appendix, we prove \hyperlink{Lemma4}{Lemma~4}.

\hypertarget{Lemma4}{}
\vspace{\baselineskip} \noindent \textsc{Lemma 4 (Restated):} Fix a setting with two agents in $N = \{1, 2\}$ and let $e \in E$. For each level~$\ell$, each valid partition record $(\mathcal{P}, \alpha)$ that is $\ell$-deficient, and each $\textsf{q} \in \mathcal{Q}$, there is $(\mathcal{P}^+, \alpha^+) \in \mathcal{R}(\mathsf{q}|\mathcal{P}, \alpha)$ that is $\lfloor \sqrt{\ell/2} \rfloor$-deficient.

\vspace{\baselineskip} \noindent \textsc{Proof:} Let $\ell$, $(\mathcal{P}, \alpha)$, and $\mathsf{q}$ satisfy the hypotheses. To complete the proof, we select an arbitrary kitchen measure profile $\mu$ from a nonempty set, then construct the desired $(\mathcal{P}^+, \alpha^+)$.

\vspace{\baselineskip} \noindent \textsc{Step 1:} {\it Introduce preliminary notation and select the kitchen measure profile $\mu$, then construct the serving $B'$ and the table $\mathcal{T}$.}

\vspace{\baselineskip} First, we introduce preliminary notation similar to that used in the proof of \hyperlink{Lemma3}{Lemma~3}. Define $\ell^+ \equiv \lfloor \sqrt{\ell/2} \rfloor$, let $s \equiv |\mathcal{P}|$ denote the number of slices in $\mathcal{P}$, and index the members of $\mathcal{P}$ by $\{A_j\}_{j \in \{1, 2, ..., s\}}$. Denote the arguments of $\mathsf{q}$ by $(1, \kappa, B, p_1)$, so that (re-indexing if necessary) $\mathsf{q}$ asks agent $1$ to cut, and let $2$ denote the other agent; such a re-indexing is without loss of generality as argued at the start of the proof of \hyperlink{Lemma3}{Lemma~3}.

Second, we construct a table of servings, each given by a row and a column, and from this table we construct $\mathcal{P}^+$. We take $\mathcal{P}$ to be the set of {\it rows}. Since $(\mathcal{P}, \alpha)$ is valid, thus we can select an arbitrary $\mu \in \mathbb{D}(\mathcal{P}, \alpha)$. We define $B' \equiv S_{\mathsf{q}|\mu_1}$, we define the set of {\it columns} $\mathcal{C} \equiv \{B', B \backslash B', C \backslash B\}$, and we define the {\it table} $\mathcal{T} \equiv \{A \cap A' | A \in \mathcal{P}, A' \in \mathcal{C} \}$.

Finally, we highlight an abuse of notation: as in the proof of \hyperlink{Lemma3}{Lemma~3}, we always use the same notation for an appraisal, its subalgebra extension, and its hyperallocation extension.

Before proceeding, we provide a brief outline. Having used $\mu$ to construct $\mathcal{T}$, the plan is to (i)~modify $\mu_1{\restriction}_\mathcal{T}$ to $\alpha^\approx_1$, so that at most one row assigns positive value to multiple entries, (ii)~modify $\alpha^\approx_1$ to $\alpha^\mathcal{T}_1$, so that if there is a unique row that assigns positive value to multiple entries then it assigns positive value to exactly two entries, (iii)~in that case, use $\alpha^\mathcal{T}_1$ to construct an associated query $\mathsf{q}^*$ whose serving argument is this unique row, (iv)~apply \hyperlink{Lemma3}{Lemma~3} to construct an $\ell^+$-deficient ultraresponse $(\mathcal{P}^*, \alpha^*)$ to $\mathsf{q}^*$, (v)~use $\alpha^*_2$ on $\mathcal{P}^*$ to construct $\alpha^\mathcal{T}_2$ on $\mathcal{T}$, so that $(\mathcal{T}, \alpha^\mathcal{T})$ ``inherits" $\ell^+$-deficiency from $(\mathcal{P}^*, \alpha^*)$, and (vi)~mix $\alpha^\mathcal{T}$ slightly with $\mu$ to ensure servings that are not slivers get value without breaking $\ell^+$-deficiency, at which point we have the desired ultraresponse $(\mathcal{P}^+, \alpha^+)$.

\vspace{\baselineskip} \noindent \textsc{Step 2:} {\it Index the table entries, then introduce cutter appraisals and split rows.}

\vspace{\baselineskip} First, we index the table entries. In particular, for each $j \in \{1, 2, ..., s\}$, we define (i)~$B'_j \equiv A_j \cap B'$, (ii)~$B''_j \equiv A_j \cap (B \backslash B')$, and (iii)~$B'''_j \equiv A_j \cap (C \backslash B)$.

Second, let us say that a function $\alpha: \mathcal{T} \to [0, 1]$ is a {\it cutter appraisal} if it satisfies the following requirements:
\begin{itemize}
\item Each row's value according to agent $1$ is distributed across its entries: for each row $j \in \{1, 2, ..., s\}$, $\alpha(B'_j) + \alpha(B''_j) + \alpha(B'''_j) = \alpha_1(A_j)$.

\item No sliver is assigned any value: for each $S \in \mathcal{T} \cap \mathcal{S}^\circ$, $\alpha(S) = 0$.

\item The ratio of the value in the first column to the value in the first two columns is precisely the target proportion: $\sum_{j=1}^s  \alpha(B'_j) = p_1  \cdot ( \sum_{j=1}^s [\alpha(B'_j) + \alpha(B''_j)] )$.
\end{itemize}
Let $\mathbb{A}$ denote the collection of cutter appraisals. Observe that a cutter appraisal need not be the restriction of a kitchen measure to $\mathcal{T}$ because it may assign zero to a serving that is not a sliver.

Finally, for each $\alpha \in \mathbb{A}$ and each $j \in \{1, 2, ..., s\}$, we say that row $j$ is {\it $\alpha$-split} if $\alpha$ assigns positive value to at least two $j$ entries: $|\{B \in \{B'_j, B''_j, B'''_j\} | \alpha(B)> 0\}| > 1$.

\vspace{\baselineskip} \noindent \textsc{Step 3:} Construct $\alpha^\mathcal{T}_1 \in \mathbb{A}$ such that (i)~there is at most one $\alpha^\mathcal{T}_1$-split row, and (ii)~if there is an $\alpha^\mathcal{T}_1$-split row, then it assigns positive value to exactly two entries.

\vspace{\baselineskip} We first handle the easy case $p_1 \in \{0, 1\}$. In this case, for each row $j$ such that $\alpha_1(A_j) > 0$, let $\alpha^\mathcal{T}_1$ assign the entire value of the row to the leftmost (as written here) member of $\{B'_j, B''_j, B'''_j\}$ that is not a sliver. It is easy to verify that there is no $\alpha^\mathcal{T}_1$-split row, and thus that $\alpha^\mathcal{T}_1$ has the desired properties.

Thus let us assume $p_1 \in (0, 1)$. There are many ways to construct the desired $\alpha^\mathcal{T}_1$; what follows is simply one concrete example. We begin with $\mu_1{\restriction}_\mathcal{T} \in \mathbb{A}$, then apply our \textsc{Row Polarization} procedure to construct a cutter appraisal $\alpha^\approx_1$ with at most one $\alpha^\approx_1$-split row, and finally modify $\alpha^\approx_1$ to reach the desired $\alpha^\mathcal{T}_1$.

Formally, the \textsc{Row Polarization} procedure begins with $\mu_1{\restriction}_\mathcal{T}  \in \mathbb{A}$ as the input to the first stage. At each stage, if the input $\alpha  \in \mathbb{A}$ is such that there is at most one $\alpha$-split row, then the procedure terminates with $\alpha^\approx_1 \equiv \alpha$; otherwise, it constructs the output $\alpha' \in \mathbb{A}$ that serves as the input to the next stage as follows.
\begin{itemize}
\item Let $j_1$ denote the smallest index $\{1, 2, ..., s\}$ of an $\alpha$-split row and let $j_2$ denote the largest such index. Since there are at least two $\alpha$-split rows, $j_1 \neq j_2$.

\item Define $S^{source}_1$ and $S^{sink}_1$ in row $j_1$ as follows:
\begin{itemize}
\item If $\alpha(B'_{j_1}) > 0$ and $\alpha(B''_{j_1}) > 0$, then $S^{source}_1 \equiv B''_{j_1}$ and $S^{sink}_1 \equiv B'_{j_1}$.

\item Else if $\alpha(B'_{j_1}) > 0$ and $\alpha(B'''_{j_1}) > 0$, then $S^{source}_1 \equiv B'''_{j_1}$ and $S^{sink}_1 \equiv B'_{j_1}$.

\item Else $\alpha(B''_{j_1}) > 0$ and $\alpha(B'''_{j_1}) > 0$, and we define $S^{source}_1 \equiv B''_{j_1}$ and $S^{sink}_1 \equiv B'''_{j_1}$.
\end{itemize}
Observe that transferring value from $S^{source}_1$ to $S^{sink}_1$ increases the ratio of the value in the first column to the value in the first two columns.

\item Define $S^{source}_2$ and $S^{sink}_2$ in row $j_2$ as follows:
\begin{itemize}
\item If $\alpha(B'_{j_2}) > 0$ and $\alpha(B''_{j_2}) > 0$, then $S^{source}_2 \equiv B'_{j_2}$ and $S^{sink}_2 \equiv B''_{j_2}$.

\item Else if $\alpha(B'_{j_2}) > 0$ and $\alpha(B'''_{j_2}) > 0$, then $S^{source}_2 \equiv B'_{j_2}$ and $S^{sink}_2 \equiv B'''_{j_2}$.

\item Else $\alpha(B''_{j_2}) > 0$ and $\alpha(B'''_{j_2}) > 0$, and we define $S^{source}_2 \equiv B'''_{j_2}$ and $S^{sink}_2 \equiv B''_{j_2}$.
\end{itemize}
Observe that transferring value from $S^{source}_2$ to $S^{sink}_2$ decreases the ratio of the value in the first column to the value in the first two columns.

\item For each value transfer rate $\lambda \in \mathbb{R}_+$, define $\alpha_\lambda: \mathcal{T} \to \mathbb{R}$ as follows: (i)~for each $S \in \mathcal{T} \backslash \{S^{source}_1, S^{sink}_1, S^{source}_2, S^{sink}_2\}$, $\alpha_\lambda(S) \equiv \alpha(S)$, (ii)~$\alpha_\lambda(S^{source}_1) \equiv \alpha(S^{source}_1) - 1$, (iii)~$\alpha_\lambda(S^{sink}_1) \equiv \alpha(S^{sink}_1) + 1$, (iv)~$\alpha_\lambda(S^{source}_2) \equiv \alpha(S^{source}_2) - \lambda$, and (v)~$\alpha_\lambda(S^{sink}_2) \equiv \alpha(S^{sink}_2) + \lambda$.

\item Let $\lambda^*$ be the unique $\lambda \in \mathbb{R}_+$ such that $\sum_{j=1}^s  \alpha_\lambda(B'_j) = p_1 \cdot (\sum_{j=1}^s [\alpha_\lambda(B'_j) + \alpha_\lambda(B''_j)])$. Since $p_1 \in (0, 1)$, it is easy to verify that $\lambda^*$ is well-defined, that $\lambda^*$ is greater than zero, and that $\lambda^*$ can be explicitly constructed in cases given by the sinks and sources; we omit the details.

\item Define $v \equiv \min \{\alpha(S^{source}_1), \frac{1}{\lambda^*} \cdot \alpha(S^{source}_2)\}$, and define $\alpha': \mathcal{T} \to [0, 1]$ as follows: (i)~for each $S \in \mathcal{T} \backslash \{S^{source}_1, S^{sink}_1, S^{source}_2, S^{sink}_2\}$, $\alpha'(S) \equiv \alpha(S)$, (ii)~$\alpha'(S^{source}_1) \equiv \alpha(S^{source}_1) - v$, (iii)~$\alpha'(S^{sink}_1) \equiv \alpha(S^{sink}_1) + v$, (iv)~$\alpha'(S^{source}_2) \equiv \alpha(S^{source}_2) - \lambda^* \cdot v$, and (v)~$\alpha'(S^{sink}_2) \equiv \alpha(S^{sink}_2) + \lambda^* \cdot v$.
\end{itemize}
Intuitively, $\lambda^*$ is the rate such that if we simultaneously and continuously transfer value (i)~from source to sink in row $j_1$ at a rate of one unit per second, and (ii)~from source to sink in row $j_2$ at a rate of $\lambda^*$ units per second, then we preserve the ratio of the value in the first column to the value in the first two columns. We do so until one of the sources is drained of all value; necessarily $\alpha'$ belongs to $\mathbb{A}$.

Observe that the \textsc{Row Polarization} procedure must terminate with $\alpha^\approx_1$. Indeed, at each stage for which the input and output differ, the output has more zero value assignments than the input, and the number of zero value assignments cannot exceed the total number of servings in the table.

Finally, we define $\alpha^\mathcal{T}_1$ by modifying $\alpha^\approx_1$ as follows: (i)~if there is a unique $\alpha^\approx_1$-split row $j^*$, and if that row assigns positive value to all three entries, then transfer $p_1 \cdot \alpha^\approx_1(B'''_{j^*})$ from $B'''_{j^*}$ to $B'_{j^*}$ and transfer $(1-p_1) \cdot \alpha^\approx_1(B'''_{j^*})$ from $B'''_{j^*}$ to $B''_{j^*}$, and (ii)~otherwise make no modification. It is easy to verify that the result has the desired properties.

\vspace{\baselineskip} \noindent \textsc{Step 4:} {\it Conclude.}

\vspace{\baselineskip} First, we use $\alpha^\mathcal{T}_1$ to construct an associated $(\mathcal{P}^*, \alpha^*)$ that is $\ell^+$-deficient. If there are no $\alpha^\mathcal{T}_1$-split rows, then we simply take $(\mathcal{P}^*, \alpha^*) \equiv (\mathcal{P}, \alpha)$. Otherwise, (i)~let $j^*$ be the index of the unique $\alpha^\mathcal{T}_1$-split row, (ii)~let $C'_{j^*}$ denote the leftmost (as written here) member of $\{B'_{j^*}, B''_{j^*}, B'''_{j^*}\}$ that $\alpha^\mathcal{T}_1$ assigns positive value and let $C''_{j^*}$ denote the other member of $\{B'_{j^*}, B''_{j^*}, B'''_{j^*}\}$ that $\alpha^\mathcal{T}_1$ assigns positive value, (iii)~define $p^*_1 \in (0, 1)$ to be the associated value ratio, $p^*_1 \equiv \frac{\alpha^\mathcal{T}_1(C'_{j^*})}{\alpha^\mathcal{T}_1(C'_{j^*})+\alpha^\mathcal{T}_1(C''_{j^*})}$, which for emphasis does not involve division by zero construction of $C'_{j^*}$ and $C''_{j^*}$, (iv)~define the query $\mathsf{q}^* \equiv (1, \kappa, A_{j^*}, p^*_1)$, and (v)~let $(\mathcal{P}^*, \alpha^*)$ be the $\ell^+$-deficient member of $\mathcal{R}(\mathsf{q}^*|\mathcal{P}, \alpha)$ promised by \hyperlink{Lemma3}{Lemma~3} and constructed in its proof. In the latter case, there is $A'_{j^*} \in \mathcal{P}^* \backslash \mathcal{P}$ such that
\begin{align*}
\alpha^*_1(A'_{j^*}) &= p^*_1 \cdot \alpha^*_1(A_{j^*})
\\ &= \Big[ \frac{\alpha^\mathcal{T}_1(C'_{j^*})}{\alpha^\mathcal{T}_1(C'_{j^*})+\alpha^\mathcal{T}_1(C''_{j^*})}\Big] \cdot [\alpha^\mathcal{T}_1(B'_{j^*})+\alpha^\mathcal{T}_1(B''_{j^*}) +\alpha^\mathcal{T}_1(B'''_{j^*})]
\\ &= \Big[ \frac{\alpha^\mathcal{T}_1(C'_{j^*})}{\alpha^\mathcal{T}_1(C'_{j^*})+\alpha^\mathcal{T}_1(C''_{j^*})} \Big] \cdot [\alpha^\mathcal{T}_1(C'_{j^*})+\alpha^\mathcal{T}_1(C''_{j^*})]
\\ &= \alpha^\mathcal{T}_1(C'_{j^*}).
\end{align*}
Define $A''_{j^*} \equiv A_{j^*} \backslash A'_{j^*}$.

Second, we define an injection $f: \mathcal{P}^* \to \mathcal{T}$ such that for each $S \in \mathcal{P}^*$, $\alpha^*_1(S) = \alpha^\mathcal{T}_1(f(S))$. First, for each $j \in \{1, 2, ..., s\}$ that is not $\alpha^\mathcal{T}_1$-split, (i)~$\alpha^*_1(A_j) = 0$ implies $f(A_j)$ is the leftmost (as written here) member of $\{B'_j, B''_j, B'''_j\}$ that is not empty, and (ii)~$\alpha^*_1(A_j) > 0$ implies $f(A_j)$ is the unique $B \in \{B'_j, B''_j, B'''_j\}$ such that $\alpha^\mathcal{T}_1(B) > 0$. Second, if there is a unique $\alpha^\mathcal{T}_1$-split row $j^*$, then define $f(A'_{j^*}) \equiv C'_{j^*}$ and $f(A''_{j^*}) \equiv C''_{j^*}$. It is straightforward to verify that $f$ satisfies the desired properties.

Third, we construct $\alpha^\mathcal{T}_2: \mathcal{T} \to [0,1]$ such that (i)~value is only assigned to $f(\mathcal{P}^*) \subseteq \mathcal{T}$, and (ii)~for each $S \in \mathcal{P}^*$, $\alpha^*_2(S) = \alpha^\mathcal{T}_2(f(S))$. First, for each $S \in \mathcal{P}^*$, define $\alpha^\mathcal{T}_2(f(S)) \equiv \alpha^*_2(S)$. Second, for each $S \in \mathcal{T} \backslash f(\mathcal{P}^*)$, define $\alpha^\mathcal{T}_2(S) \equiv 0$.

Fourth, for each $X \in \mathcal{X}_{\ell^+| \mathcal{P}^*}$, let $X^\mathcal{T} \in \mathcal{X}_{\ell^+|\mathcal{T}}$ be the associated hyperallocation with $r(X^\mathcal{T}) = r(X)$ such that for each $r \in \{1, 2, ..., r(X^\mathcal{T})\}$, $X^\mathcal{T}_r \equiv \cup \{f(S) | S \in \mathcal{P}^*, S \subseteq X_r\}$. In other words, $X^\mathcal{T}_r$ assigns each serving $f(S) \in f(\mathcal{P}^*)$ to the agent to whom $X_r$ assigns $S$, which is well-defined because $f$ is an injection, and then assigns all servings in $\mathcal{T} \backslash f(\mathcal{P}^*)$ to agent~$2$. Observe that by construction, (i)~for each $i \in N$ and each $X \in \mathcal{X}_{\ell^+|\mathcal{P}^*}$, we have $\alpha^\mathcal{T}_i(X^\mathcal{T}) = \alpha^*_i(X)$, and (ii)~for each $X' \in \mathcal{X}_{\ell^+|\mathcal{T}}$, there is $X \in \mathcal{X}_{\ell^+|\mathcal{P}^*}$ such that $X'$ is equivalent to $X^\mathcal{T}$ according to both $\alpha^\mathcal{T}_1$ and $\alpha^\mathcal{T}_2$. The latter holds because both $\alpha^\mathcal{T}_1$ and $\alpha^\mathcal{T}_2$ assign all value to $f(\mathcal{P}^*)$.

To conclude, we construct the desired ultraresponse $(\mathcal{P}^+, \alpha^+)$. We begin by modifying $(\alpha^\mathcal{T}_1, \alpha^\mathcal{T}_2)$ to assign positive value to servings that are not slivers. First, since $(\mathcal{P}^*, \alpha^*)$ is $\ell^+$-deficient, thus for each $X \in \mathcal{X}_{\ell^+| \mathcal{P}^*}$, either $\alpha^*_1(X) < e_1$ or $\alpha^*_2(X) > e_1$, so by the previous paragraph, the associated {\it deficit for $X^\mathcal{T}$}
\begin{align*}
\delta(X^\mathcal{T}) \equiv \max \{\max\{e_1 - \alpha^\mathcal{T}_1(X^\mathcal{T}), 0\}, \max\{e_2 - (1-\alpha^\mathcal{T}_2(X^\mathcal{T})),0\}\}
\end{align*}
is positive. By the previous paragraph again, the {\it minimum deficit} $\min_{X' \in \mathcal{X}_{\ell^+| \mathcal{T}}} \delta(X')$ is positive. Thus there is $\varepsilon > 0$ such that if we mix $(\alpha^\mathcal{T}_1, \alpha^\mathcal{T}_2)$ with the original measure profile $\mu$ by putting weight $\varepsilon$ on the latter, then the associated minimum deficit remains positive. If we then discard all empty members of $\mathcal{T}$ to construct $\mathcal{P^+}$, and define $(\alpha^+_1, \alpha^+_2)$ by taking the restriction of the preceding mixture to $\mathcal{P^+}$, the result is an $\ell^+$-deficient ultraresponse $(\mathcal{P}^+, \alpha^+) \in \mathcal{R}(\mathsf{q}|\mathcal{P}, \alpha)$, as desired.~$\blacksquare$

\hypertarget{Appendix3}{}
\setcounter{secnumdepth}{0}
\section{Appendix 3}

In this appendix, we prove \hyperlink{Proposition1}{Proposition~1}.

\vspace{\baselineskip} \noindent \textsc{Proposition 1 (Restated):} Fix a setting. For each $e \in E$ and each precision level $\mathsf{p} \in \mathbb{N}$, $\mathsf{P}(e) > \mathsf{p}$ implies $\mathsf{cost}(e) > \lfloor \log_2 \log_2 2 \mathsf{p} \rfloor$.

\vspace{\baselineskip} \noindent \textsc{Proof:} We first establish the proposition for two-agent settings, then conclude.

\vspace{\baselineskip} \noindent \textsc{Step 1:} {\it Prove the proposition for two-agent settings.}

\vspace{\baselineskip} Fix a two-agent setting, let $\mathsf{p} \in \mathbb{N}$, and let $e \in E$ be such that $\mathsf{P}(e) > \mathsf{p}$. Since $\mathsf{P}(e) > \mathsf{p}$, thus there is $i \in N$ such that $\mathsf{P}_i(e) > \mathsf{p}$; denote the other agent by $j$. Finally, let $\pi \in \Pi_e$.

We begin by introducing some preliminary concepts. First, we say that each $h \in H$ is {\it $0$-deficient}. Moreover, for each $h \in H$ and each level $\ell \in \mathbb{N}_0$, we say that $h$ is {\it $\ell$-deficient} if there is a valid $\ell$-deficient partition record $(\mathcal{P}, \alpha)$ such that $\mathbb{D}(h) \supseteq \mathbb{D}(\mathcal{P}, \alpha)$. Second, define the initial level $\ell_0 \equiv \mathsf{p}$, and for each query count $c \in \mathbb{N}_0$ define the level $\ell_{c+1} \equiv \lfloor \sqrt{\ell_c/2} \rfloor$. Finally, for each $c \in \mathbb{N}_0$ and each $\mu \in \mathbb{D}$, define $h_c(\mu) \in H \cup \{\mathsf{error}\}$ as follows: (i)~if $c \leq \mathsf{cost}(\pi|\mu)$, then $h_c(\mu)$ is the history in the play determined by $\mu$ and $\pi$ whose query count is $c$, and (ii)~otherwise $h_c(\mu) \equiv \mathsf{error}$.

We claim that for each $c \in \mathbb{N}_0$ such that $\ell_c \geq 1$, there is $\mu \in \mathbb{D}$ such that for each $c' \in \{0, 1, ..., c+1\}$, $h_{c'}(\mu) \in H$ and $h_{c'}(\mu)$ is $\ell_{c'}$-deficient. We proceed by induction. For the base step, since $\mathsf{P}_i(e) > \mathsf{p}$, thus $e_i$ cannot be written as a fraction whose denominator is at most $\ell_0 = \mathsf{p}$, so the unique valid partition record for partition $\{C\}$ is $\ell_0$-deficient, so by \hyperlink{Lemma4}{Lemma~4} there is an $\ell_1$-deficient ultraresponse to the initial query of $\pi$ associated with a $\mu$ meeting the desired requirements; thus the claim holds for $c=0$. For the inductive step, let $c \in \mathbb{N}_0$ be such that the claim holds with measure profile $\mu$. If $\ell_{c+1} < 1$ then we are done, so assume $\ell_{c+1} \geq 1$. By the inductive hypothesis, $h_{c+1}(\mu)$ is $\ell_{c+1}$-deficient, so there is a valid $\ell_{c+1}$-deficient partition record $(\mathcal{P}, \alpha)$ such that $\mathbb{D}(h_{c+1}(\mu)) \supseteq \mathbb{D}(\mathcal{P}, \alpha)$. Since $\ell_{c+1} \geq 1$, there is no $X \in \mathcal{X}$ such that for each $\mu \in \mathbb{D}(\mathcal{P}, \alpha)$, $X$ is $(e|\mu)$-proportional; thus since $\pi \in \Pi_e$, necessarily $\pi(h_{c+1}(\mu))$ is a query $\mathsf{q}$. By \hyperlink{Lemma4}{Lemma~4}, there is $\mu' \in \mathbb{D}(\mathcal{P}, \alpha)$ such that the ultraresponse $\mathsf{r}(\mathsf{q}|\mu', \mathcal{P}, \alpha)$ is $\lfloor \sqrt{\ell_{c+1}/2} \rfloor$-deficient, or equivalently $\ell_{c+2}$-deficient. Since $\mu' \in \mathbb{D}(\mathcal{P}, \alpha) \subseteq \mathbb{D}(h_{c+1}(\mu))$, thus $\mu$ and $\mu'$ share the same $c+1$ queries and responses. Altogether, then, the claim holds for $c+1$ with measure profile $\mu'$, so as $c \in \mathbb{N}_0$ was arbitrary we are done.

To conclude, define $c^* \equiv \max \{c \in \mathbb{N}_0 | \mathsf{p} \geq 2^{2^c-1}\}$, which is well-defined since $\mathsf{p} \geq 1$. We claim that for each $c \in \{0, 1, ..., c^*\}$, $\ell_c \geq 2^{2^{c^*-c}-1}$. We proceed by induction. For the base step, $\ell_0 = \mathsf{p} \geq 2^{2^{c^*}-1}$, as desired. For the inductive step, let $c \in \{0, 1, ..., c^*-1\}$ be such that $\ell_c \geq 2^{2^{c^*-c}-1}$. Then $\ell_{c+1} = \lfloor \sqrt{\ell_c/2} \rfloor \geq \lfloor \sqrt{2^{2^{c^*-c}-1}/2} \rfloor = \lfloor \sqrt{2^{2^{c^*-c}-2}} \rfloor = \lfloor 2^{2^{c^*-(c+1)}-1} \rfloor = 2^{2^{c^*-(c+1)}-1}$, so the claim holds for $c+1$. Since $c \in \{0, 1, ..., c^*-1\}$ was arbitrary, we are done. In particular, we have that $\ell_{c^*} \geq 1$, so by the previous paragraph there is $\mu \in \mathbb{D}$ such that $h_{c^*+1}(\mu) \in H$, so $\mathsf{cost}(\pi|\mu) > c^*$, so $\mathsf{cost}(\pi) > c^*$. Since $\pi \in \Pi_e$ was arbitrary, thus $\mathsf{cost}(e) > c^*$. Finally, by definition of $c^*$ we have that $c^* \geq \lfloor \log_2 \log_2 2\mathsf{p} \rfloor$. Altogether, then, we have $\mathsf{cost}(e) > \lfloor \log_2 \log_2 2\mathsf{p} \rfloor$, as desired.

\vspace{\baselineskip} \noindent \textsc{Step 2:} {\it Conclude.}

\vspace{\baselineskip} Assume, by way of contradiction, there is a setting $(C, \mathbb{K}, N)$ for which the proposition does not hold: there are $\mathsf{p} \in \mathbb{N}$, $e \in E$ with $\mathsf{P}(e) > \mathsf{p}$, and $\pi \in \Pi_e$ such that for each $\mu \in \mathbb{D}$, $\mathsf{cost}(\pi|\mu) \leq \lfloor \log_2 \log_2 2\mathsf{p} \rfloor$. For each setting with one agent, each entitlement profile's precision index is $1$ and thus the proposition holds vacuously, so $n \neq 1$. Moreover, for each setting with two agents, the proposition holds by Step~1, so $n \neq 2$. Altogether, then, $n > 2$. We refer to $(C, \mathbb{K}, N)$ as the {\it large setting}. Since $\mathsf{P}(e) > \mathsf{p}$, thus there is $i \in N$ such that $\mathsf{P}_i(e) > \mathsf{p}$. Select $j \in N \backslash \{i\}$ and define $N' \equiv \{i, j\}$. We refer to $(C, \mathbb{K}, N')$ with $|N'| = 2$ as the {\it small setting}.

Since our notation suppresses the setting while the following argument involves two settings, we simply use~$'$ to distinguish objects for the small setting from the associated objects from the large setting---for example, writing $E$ for the large setting and $E'$ for the small setting.

To complete this step, we prove that the proposition does not hold for the small setting, which contradicts Step~1. More precisely, define $e' \in E'$ by $(e'_i, e'_j) \equiv (e_i, 1 - e_i)$. By construction, $\mathsf{P}'(e') \geq \mathsf{P}'_i(e') = \mathsf{P}_i(e) > \mathsf{p}$. We claim that $\mathsf{cost}'(e') \leq \lfloor \log_2 \log_2 2\mathsf{p} \rfloor$, which we establish by exhibiting a protocol $\pi' \in \Pi'_{e'}$ such that $\mathsf{cost}(\pi') \leq \lfloor \log_2 \log_2 2\mathsf{p} \rfloor$. Intuitively, the protocol $\pi'$ that we exhibit imitates $\pi$ in a suitable sense.

We begin by introducing some objects that we use to construct $\pi'$. First, select a {\it guess function} $\mathsf{g}: 2^{\mathbb{D}'} \backslash \{\emptyset\} \to \mathbb{D}'$ that selects from each nonempty domain $\mathbb{D}^* \subseteq \mathbb{D}'$ a guess $\mathsf{g}(\mathbb{D}^*) \in \mathbb{D}^*$. Second, define $\mathbb{D}^i \equiv \{ \mu \in \mathbb{D} | \text{ for each pair } k, k' \in N \backslash \{i\}, \mu_k = \mu_{k'}\}$, and define the {\it extension function} $\mathcal{E}: \mathbb{D}' \to \mathbb{D}^i$ be such that for each $\mu' \in \mathbb{D}'$, $\mathcal{E}(\mu')$ is the $\mu \in \mathbb{D}^i$ such that (i)~$\mu_i = \mu'_i$, and (ii)~for each $k \in N \backslash \{i\}$, $\mu_k = \mu'_j$. Finally, for each $c \in \mathbb{N}_0$ and each $\mu \in \mathbb{D}$, we define $h_c(\pi|\mu) \in H \cup \{\mathsf{error}\}$ as follows: (i)~if $c \leq \mathsf{cost}(\pi|\mu)$, then $h_c(\pi|\mu)$ is the history in the play determined by $\mu$ and $\pi$ whose query count is $c$, and (ii)~otherwise $h_c(\pi|\mu) \equiv \mathsf{error}$.

Next, we construct $\pi'$. In particular, for each $h' \in H'$, (i)~let $c$ denote the query count at $h'$, (ii)~define the guess $\mathsf{g}' \equiv \mathsf{g}(\mathbb{D}'(h'))$, (iii)~define the guess extension $\mathsf{g} \equiv \mathcal{E}(\mathsf{g}')$, and (iv)~define $\pi'(h') \in \mathcal{Q}' \cup \mathcal{X}'$ in three cases as follows.
\begin{itemize}
\item If $c < \mathsf{cost}(\pi|\mathsf{g})$, then $\pi(h_c(\pi|\mathsf{g})) \in \mathcal{Q}$. If the cutter of this query belongs to $N'$, then define $\pi'(h')$ to be the same query; otherwise define $\pi'(h')$ to be query formed by taking this query and then making $j$ the cutter.

\item If $c = \mathsf{cost}(\pi|\mathsf{g})$, then define $X \equiv \mathsf{out}(\pi|\mathsf{g})$, let $\pi'(h') \in \mathcal{X}'$ be defined by $X'_i = X_i$ and $X'_j \equiv C \backslash X'_i$, and define $\pi'(h') \equiv X'$. Since $\pi \in \Pi_e$, thus $X$ is $(e|\mathsf{g})$-proportional, from which it follows that $X'$ is $(e'|\mathsf{g}')$-proportional.

\item If $c > \mathsf{cost}(\pi|\mathsf{g})$, then $h'$ will ultimately be off-path. For concreteness, set $\pi'(h') \in \mathcal{X}'$ to be the allocation at which agent $i$ receives the entire cake.
\end{itemize}
Observe that since $\pi'$ assigns the same action to any two histories with the same chronicle, as each history's guess only depends on its chronicle, thus $\pi'$ is indeed a protocol. For each $c \in \mathbb{N}_0$ and each $\mu' \in \mathbb{D}'$, we define $h'_c(\pi'|\mu') \in H' \cup \{\mathsf{error}\}$ analogously to the associated notation for $\pi$.

\vspace{\baselineskip} \noindent {\it Claim.} Fix an arbitrary {\it true} measure profile $\mathsf{t}' \in \mathbb{D}'$. Let us write $\mathsf{t} \equiv \mathcal{E}(\mathsf{t}')$, and for each $c \in \mathbb{N}_0$ such that $h'_c \in H'$, let us write $h'_c \equiv h'_c(\pi'|\mathsf{t}')$, $\mathsf{g}'_c \equiv \mathsf{g}(\mathbb{D}'(h'_c))$, $\mathsf{g}_c \equiv \mathcal{E}(\mathsf{g}'_c)$, $h^\mathsf{t}_c \equiv h_c(\pi|\mathsf{t})$, and $h^\mathsf{g}_c \equiv h_c(\pi|\mathsf{g}_c)$. We claim that for each $c \in \mathbb{N}_0$ such that $h'_c \in H'$, we have
\begin{itemize}
\item $h_c^\mathsf{t} \in H$ and $h_c^\mathsf{g} \in H$,

\item $\mathbb{D}(h_c^\mathsf{t}) = \mathbb{D}(h_c^\mathsf{g})$, and

\item $\mathbb{D}(h_c^\mathsf{t}) \cap \mathbb{D}^i = \mathcal{E}(\mathbb{D}'(h'_c))$.
\end{itemize}

\vspace{\baselineskip} \noindent {\it Proof of Claim.} We prove the claim by induction on the query count, and the base step $c=0$ is trivial; thus let us make the inductive hypothesis that the claim holds for $c \in \mathbb{N}_0$. If $h'_{c+1} = \mathsf{error}$ then we are done; thus let us assume $h'_{c+1} \in H'$, in which case $\pi'(h'_c) \in \mathcal{Q}'$. Then by construction of $\pi'$, $h^\mathsf{g}_c \in H$ and $\pi(h^\mathsf{g}_c) \in \mathcal{Q}$, so $h^\mathsf{g}_{c+1} \in H$. Moreover, (i)~since $\mathsf{t}'$ and $\mathsf{g}'_c$ both belong to $\mathbb{D}'(h'_c)$, thus by the inductive hypothesis $\mathsf{t}$ and $\mathsf{g}_c$ both belong to $\mathbb{D}(h^\mathsf{t}_c)$, and (ii)~$\mathsf{g}_c$ belongs to $\mathbb{D}(h^\mathsf{g}_c)$; thus necessarily $\mathbb{D}(h^\mathsf{t}_c)=\mathbb{D}(h^\mathsf{g}_c)$.\footnote{Indeed, this follows from two observations: (i)~if the domains of two histories intersect, then they are nested, and (ii)~if the domains of two histories are strictly nested, then their query counts are not the same.} Altogether, then, $\pi(h^\mathsf{t}_c) = \pi(h^\mathsf{g}_c) \in \mathcal{Q}$, so $h_{c+1}^\mathsf{t} \in H$. This establishes the first conclusion, from which the second conclusion follows by definition of $\mathsf{g}_{c+1}$. To conclude, define $\mathsf{q} \equiv \pi(h^\mathsf{t}_c) = \pi(h^\mathsf{g}_c)$ and define $\mathsf{q}' \equiv \pi'(h'_c)$. Using the definition of $\pi'$, it is straightforward to verify that whether or not the cutter of this query belongs to $N'$, we have $\mathcal{E}(\mathsf{r}'(\mathsf{q}' | \mathsf{t}')) = \mathsf{r}(\mathsf{q} | \mathsf{t}) \cap \mathbb{D}^i$, which together with the inductive hypothesis directly implies the third conclusion. Altogether, then, the claim holds for $c+1$, so as $c \in \mathbb{N}_0$ was arbitrary we are done.

\vspace{\baselineskip} \noindent {\it Proof from Claim.} To conclude, let $\mathsf{t}' \in \mathbb{D}'$. Define $\mathsf{t} \equiv \mathcal{E}(\mathsf{t}')$, define $X' \equiv \mathsf{out}(\pi'|\mathsf{t}')$ and define $X \equiv \mathsf{out}(\pi|\mathsf{t})$. We claim that $\mathsf{cost}'(\pi'|\mathsf{t}') \leq \lfloor \log_2 \log_2 2\mathsf{p} \rfloor$ and $X'$ is $(e'|\mathsf{t}')$-proportional. Indeed, by our hypotheses about $\pi$, we have that $\mathsf{cost}(\pi|\mathsf{t}) \leq \lfloor \log_2 \log_2 2\mathsf{p} \rfloor$, $X \in \mathcal{X}$, and $X$ is $(e|\mathsf{t})$-proportional. By our Claim and the definition of $\pi'$, the sequence of actions in the play determined by $\pi'$ and $\mathsf{t}'$ is formed from the sequence of actions in the play determined by $\pi$ and $\mathsf{t}$ by (i)~taking any query whose cutter does not belong to $N'$ and making its cutter $j$, and (ii)~replacing $X$ with the allocation $X' \in \mathcal{X}'$ such that $X'_i = X_i$. Then $\mathsf{cost}'(\pi'|\mathsf{t}') = \mathsf{cost}(\pi|\mathsf{t}) \leq \lfloor \log_2 \log_2 2\mathsf{p} \rfloor$. Moreover, since $X$ is $(e|\mathsf{t})$-proportional, thus $\mathsf{t}'_i(X_i) \geq e_i = e'_i$; it directly follows that $X'$ is $(e'|\mathsf{t}')$-proportional. Since $\mathsf{t}' \in \mathbb{D}'$ was arbitrary, thus $\pi' \in \Pi'_{e'}$ and $\mathsf{cost}'(\pi') \leq \lfloor \log_2 \log_2 2\mathsf{p} \rfloor$, so $\mathsf{cost}'(e') \leq \lfloor \log_2 \log_2 2\mathsf{p} \rfloor$. But then the small setting is a two-agent setting for which the proposition does not hold, contradicting Step~1.~$\blacksquare$

\hypertarget{Appendix4}{}
\setcounter{secnumdepth}{0}
\section{Appendix 4}

In this appendix, we prove \hyperlink{Theorem2}{Theorem~2}.

To begin, the following lemma gathers two slight extensions of exercises in \cite{Savage1954}; the proof is omitted.

\hyperlink{Lemma5}{}
\vspace{\baselineskip} \noindent \textsc{Lemma 5:} Fix a kitchen and a qualitative probability. We have the following.
\begin{itemize}
\item For each $A \in \mathcal{S}$ and each pair $B, B' \in \mathcal{S}$ such that $B, B' \subseteq A$, we have $B \succsim B'$ if and only if $(A \backslash B') \succsim (A \backslash B)$.

\item For each four $A, A', B, B' \in \mathcal{S}$, if (i) $A \cap A' = \emptyset$, (ii)~$A \succsim B$, and (iii)~$A' \succsim B'$, then $(A \cup A') \succsim (B \cup B')$. If moreover $A \succ B$, then $(A \cup A') \succ (B \cup B')$.
\end{itemize}

\vspace{\baselineskip} We first establish that under the classic qualitative probability axioms, knife continuity is stronger than tightness.

\hyperlink{Proposition3}{}
\vspace{\baselineskip} \noindent \textsc{Proposition 3:} Fix a kitchen. If a qualitative probability satisfies {\it sliver nullity} and {\it knife continuity}, then it satisfies {\it tightness}. Moreover, for each $\kappa \in \mathbb{K}$ and each pair $C^*, A \in \mathcal{S}$ such that $C^* \succsim A$, there is $t \in [0, 1]$ such that $(C^* \cap \kappa_t) \sim A$.

\vspace{\baselineskip} \noindent \textsc{Proof:} Let $\succsim$ satisfy the hypotheses and let $\kappa \in \mathbb{K}$.

\vspace{\baselineskip} \noindent \textsc{Step 1:} For each $A \in \mathcal{S}$, there is $t \in [0, 1]$ such that $\kappa_t \sim A$.

\vspace{\baselineskip} Let $A \in \mathcal{S}$, define $T^+ \equiv \{ t \in [0, 1] | \kappa_t \succsim A \}$, and define $T^- \equiv \{ t \in [0, 1] | A \succsim \kappa_t \}$. By {\it sliver nullity} we have that $1 \in T^+$ and $0 \in T^-$, by {\it knife continuity} we have that $T^+$ and $T^-$ are closed, and by {\it completeness} we have that $T^+ \cup T^- = [0, 1]$, so since $[0, 1]$ is connected there is $t \in T^+ \cap T^-$, and necessarily $\kappa_t \sim A$, as desired.

\vspace{\baselineskip} \noindent \textsc{Step 2:} For each pair $C^*, A \in \mathcal{S}$ such that $C^* \succsim A$, there is $t \in [0, 1]$ such that $(C^* \cap \kappa_t) \sim A$.

\vspace{\baselineskip} Let $C^*, A \in \mathcal{S}$ such that $C^* \succsim A$. To ease notation, for each $x \in [0, 1]$, we define $\kappa^*_x \equiv (C^* \cap \kappa_x)$. We use $x$ to index cuts of $C^*$ and~$y$ to index cuts of $C$; thus we write $\{ \kappa^*_x \}_{x \in [0, 1]}$ and $\{\kappa_y\}_{y \in [0, 1]}$.

To begin, define $x^\leftarrow \equiv \inf \{ x \in [0, 1] | \kappa^*_x \succsim A \}$ and $x^\rightarrow \equiv \sup \{ x \in [0, 1] | A \succsim \kappa^*_x \}$; by {\it knife continuity}, {\it monotonicity}, and {\it separability}, both $x^\leftarrow$ and $x^\rightarrow$ are well-defined. By {\it completeness}, we have $x^\rightarrow \geq x^\leftarrow$, and if $x^\rightarrow > x^\leftarrow$ then for $x \equiv \frac{x^\leftarrow + x^\rightarrow}{2}$ we have $\kappa^*_x \sim A$ and we are done; thus let us assume $x^\rightarrow = x^\leftarrow$. Define $x^\circ \equiv x^\rightarrow = x^\leftarrow$, define $\kappa^*_\circ \equiv \kappa^*_{x^\circ}$, and define $\kappa_\circ \equiv \kappa_{x^\circ}$.

First, we claim that $A \succsim \kappa^*_\circ$. Indeed, assume by way of contradiction that $\kappa^*_\circ \succ A$ and define $B \equiv (\kappa_\circ \backslash C^*)$. Since $\kappa^*_\circ = (\kappa_\circ \cap C^*) \subseteq (\kappa_\circ \cap C^*) \cup (C \backslash \kappa_\circ) =  (C \backslash B)$, thus by {\it monotonicity} we have $(C \backslash B) \succsim \kappa^*_\circ \succ A$. By Step~1, there are $y(A), y(C \backslash B) \in [0, 1]$ such that $\kappa_{y(A)} \sim A$ and $\kappa_{y(C \backslash B)} \sim (C \backslash B)$, and moreover by {\it monotonicity} we have $y(C \backslash B) > y(A)$. Then $\kappa_{y(A)}$ and $(C \backslash \kappa_{y(C \backslash B)})$ are disjoint, and by \hyperlink{Lemma5}{Lemma~5} the former is equivalent to $A$ and the latter is equivalent to $B$; define $A \oplus B$ to be their union $\kappa_{y(A)} \cup (C \backslash \kappa_{y(C \backslash B)})$. For each $x \in [0, x^\circ)$, we have $x < x^\leftarrow$ and thus $\kappa^*_\circ \succ A \succ \kappa^*_x$, so by \hyperlink{Lemma5}{Lemma~5} we have $\kappa_\circ = \kappa^*_\circ \cup B \succ A \oplus B \succ \kappa^*_x \cup B \succsim \kappa_x$. But then by {\it monotonicity} we have that $\{t \in [0, 1] | A \oplus B \succsim \kappa_t\} = [0, x^\circ)$, so this set is not closed, contradicting {\it knife continuity}.

To conclude, we first claim that $\kappa^*_\circ \succsim A$. Indeed, the argument is symmetric: we omit the details, but remark that if for each $t \in [0, 1]$ we define $\kappa'_t \equiv (C \backslash \kappa_{1-t})$, then we can use the same argument using $\kappa'$ instead of $\kappa$. Altogether, then, $C^* \cap \kappa_{x^\circ} = \kappa^*_\circ \sim A$, as desired.

\vspace{\baselineskip} \noindent \textsc{Step 3:} Conclude.

\vspace{\baselineskip} We claim that for each pair $A, B \in \mathcal{S}$, if for each $A' \in \mathcal{S}$ such that $A' \succ \emptyset$ and $A \cap A' = \emptyset$ we have $A \cup A' \succsim B$, then $A \succsim B$. Indeed, let $A$ and $B$ satisfy the hypothesis. By Step~1, there are $t, t' \in [0, 1]$ such that $\kappa_t \sim A$ and $\kappa_{t'} \sim B$; thus we can define $t_A \equiv \sup \{t \in [0, 1] | A \sim \kappa_t \}$ and $t_B \equiv \inf \{t \in [0, 1] | B \sim \kappa_t\}$. By {\it knife continuity}, $\kappa_{t_A} \sim A$ and $\kappa_{t_B} \sim B$. Assume, by way of contradiction, that $t_B > t_A$, and define $t^* \equiv \frac{t_A+t_B}{2}$. By definition of $t_A$ and $t_B$ and {\it monotonicity}, we have $B \succ \kappa_{t^*} \succ A \sim \kappa_{t_A}$; thus by {\it separability} and {\it monotonicity} we have $(C \backslash A) \succsim (\kappa_{t^*} \backslash \kappa_{t_A}) \succ \emptyset$. By Step~2, there is $t \in [0, 1]$ such that for $A' \equiv (C \backslash A) \cap \kappa_t$, we have $A' \sim (\kappa_{t^*} \backslash \kappa_{t_A}) \succ \emptyset$. By \hyperlink{Lemma5}{Lemma~5} we have $A \cup A' \sim \kappa_{t_A} \cup (\kappa_{t^*} \backslash \kappa_{t_A}) = \kappa_{t^*}$, and by hypothesis we have $A \cup A' \succsim B$. But then $\kappa_{t^*} \succsim B$, contradicting $B \succ \kappa_{t^*}$. Altogether, then, $t_A \geq t_B$, so by {\it monotonicity} we have $A \sim \kappa_{t_A} \succsim \kappa_{t_B} \sim B$, as desired.

To conclude, {\it tightness} follows directly from the claim.~$\blacksquare$

\vspace{\baselineskip} The theorem follows easily from the proposition.

\vspace{\baselineskip} \noindent \textsc{Theorem 2 (Restated):} Fix a kitchen. A qualitative probability $\succsim$ satisfies {\it fineness}, {\it sliver nullity}, and {\it knife continuity} if and only if there is a kitchen measure $\mu_0$ such that for each pair $A, B \in \mathcal{S}$, we have $A \succsim B$ if and only if $\mu_0(A) \geq \mu_0(B)$. In this case, $\mu_0$ is the unique such kitchen measure.

\vspace{\baselineskip} \noindent \textsc{Proof:} It is straightforward to verify that if $\succsim$ has a kitchen measure representation, then it satisfies the axioms; we omit the argument. Thus let us assume that $\succsim$ satisfies {\it fineness}, {\it sliver nullity}, and {\it knife continuity}. By \hyperlink{Proposition3}{Proposition~3}, $\succsim$ satisfies {\it tightness}, so by \hyperlink{TheoremWM}{Theorem~WM} it has a unique probability measure representation $\mu_0$ whose range $R \equiv \{\mu_0(A) | A \in \mathcal{S}\}$ is a dense subset of the unit interval. Let $\kappa \in \mathbb{K}$ and let $f:[0, 1] \to [0, 1]$ be such that for each $t \in [0, 1]$, $f(t) = \mu_0(\kappa_t)$. For each $v \in R$, there is $A_v \in \mathcal{S}$ such that $\mu_0(A_v) = v$, so by \hyperlink{Proposition3}{Proposition~3} there is $t \in [0, 1]$ such that $\kappa_t = (C \cap \kappa_t) \sim A_v$ and thus $\mu_0(\kappa_t) = v$; thus $v \in f([0, 1])$. Since $v \in R$ was arbitrary, thus $R \subseteq f([0, 1])$, so $R = f([0, 1])$. Then the range of $f$ is a dense subset of the unit interval, and moreover by {\it monotonicity} $f$ is non-decreasing, so by a classic result\footnote{In particular, if $f:[0,1] \to [0,1]$ has dense range and is non-decreasing, then it is continuous. This result is sometimes called Froda's Theorem; see Theorem~4.30 and its corollary in \cite{Rudin1976}.} we have that $f$ is continuous. Since $f(0) = 0$ and $f(1) = 1$, thus $0 \in R$ and $1 \in R$, so by the Intermediate Value Theorem we have that $R = f([0, 1]) = [0, 1]$. From here, it follows directly from {\it sliver nullity} and \hyperlink{Proposition3}{Proposition~3} that $\mu_0$ is a kitchen measure.~$\blacksquare$

\hypertarget{Appendix5}{}
\setcounter{secnumdepth}{0}
\section{Appendix 5}

In this appendix, we prove \hyperlink{Theorem3}{Theorem~3}.

\vspace{\baselineskip} \noindent \textsc{Theorem 3 (Restated):} Fix a setting and let $e \in E$. If all entitlements are rational, then the game has a value and this value is achieved in a Nash equilibrium. If there is an irrational entitlement, then the game's value is $-\infty$ but the game has no Nash equilibrium.

\vspace{\baselineskip} \noindent \textsc{Proof:} Fix a setting and let $e \in E$.

First, we associate each protocol with a mediator strategy in the adversary game. Indeed, select an arbitrary allocation $X \in \mathcal{X}$. For each protocol $\pi \in \Pi$, let $\sigma_\pi \in \Sigma_m$ denote the following strategy: at each history $h$ in the adversary game, (i)~if $\mathbb{D}(h) \neq \emptyset$, then let $\sigma_\pi(h)$ denote the action taken at the information set in the division game whose chronicle is $h$, and (ii)~if $\mathbb{D}(h) = \emptyset$, then $\sigma_\pi(h) = X$.

Second, we associate each mediator strategy in the adversary game with a protocol. Indeed, for each $\sigma \in \Sigma_m$, let $\pi_\sigma \in \Pi$ denote the following protocol: at each information set in the division game, let $h$ denote the associated chronicle and select the action $\sigma(h)$.

Finally, we associate each measure profile with an adversary strategy. Indeed, for each measure profile $\mu \in \mathbb{D}$, let $\sigma_\mu \in \Sigma_a$ denote the following strategy: after each query $\mathsf{q}$, select $\mathsf{r}(\mathsf{q}|\mu)$.

\vspace{\baselineskip} \noindent \textsc{Case 1:} {\it All entitlements are rational.} Define $c \equiv \mathsf{cost}(e)$. By definition, there is an optimal protocol $\pi^*$ such that $\mathsf{cost}(\pi^*) = c$. Moreover, since there is a bounded protocol \citep{Steinhaus1948}, thus $c \in \mathbb{N}_0$, so there is $\mu^* \in \mathbb{D}$ such that $\mathsf{cost}(\pi^*|\mu^*) = c$. If $c=0$, then it is straightforward to verify that any strategy profile where the mediator plays $\pi^*$ is a Nash equilibrium and $0$ is the value of the adversary game; thus let us assume $c>0$.

First, consider the {\it modified game} which modifies only the payoffs as follows: (i)~if the mediator receives payoff greater than $-c$ in the adversary game, then he receives $1$ and the adversary receives $-1$ in the modified game, and (ii)~otherwise, the mediator receives $-1$ and the adversary receives $1$ in the modified game. Second, consider the {\it truncated modified game}, which immediately ends the game after $c$ queries with $-1$ for the mediator and $1$ for the adversary. By Zermelo's theorem \citep{Zermelo1913}, one of the players has a winning strategy in the truncated modified game, and it cannot be the mediator, as otherwise this yields a protocol whose cost is less than $c$, contradicting that $c$ is the cost of an optimal protocol. Thus the adversary has a winning strategy $\sigma_a$ in the truncated modified game, from which it is straightforward to verify that (i)~$\pi^*$ and any extension of $\sigma_a$ together form a Nash equilibrium of the adversary game, (ii)~the mediator's payoff at that Nash equilibrium is $-c$, and (iii)~$-c$ is the value of the adversary game.

\vspace{\baselineskip} \noindent \textsc{Case 2:} {\it There is an irrational entitlement.} First, we claim that the adversary game has no Nash equilibrium. Indeed, assume by way of contradiction that $(\sigma_m, \sigma_a)$ is a Nash equilibrium. We must have $\mathbb{P}(\sigma_m, \sigma_a) = -\infty$, else since $\pi_{\sigma_m}$ is not bounded (\hyperlink{Corollary1}{Corollary~1}), there is $\mu \in \mathbb{D}$ such that $\mathsf{cost}(\pi_{\sigma_m}|\mu) > -\mathbb{P}(\sigma_m, \sigma_a)$, so $\mathbb{P}(\sigma_m, \sigma_\mu) = -\mathsf{cost}(\pi_{\sigma_m}|\mu) < \mathbb{P}(\sigma_m, \sigma_a)$, contradicting that $(\sigma_m, \sigma_a)$ is a Nash equilibrium. But then since there is a finite protocol $\pi$ \citep{Barbanel1995}, necessarily $\mathbb{P}(\sigma_\pi, \sigma_a) > -\infty = \mathbb{P}(\sigma_m, \sigma_a)$, contradicting that $(\sigma_m, \sigma_a)$ is a Nash equilibrium.

Second, we claim that $\sup_{\sigma_m \in \Sigma_m} \inf_{\sigma_a \in \Sigma_a} \mathbb{P}(\sigma_m, \sigma_a) = -\infty$. Indeed, for each $\sigma_m \in \Sigma_m$ and each $c \in \mathbb{N}_0$, since $\pi_{\sigma_m}$ is not bounded (\hyperlink{Corollary1}{Corollary~1}) there is $\mu \in \mathbb{D}$ such that $-\mathbb{P}(\sigma_m, \sigma_\mu) = \mathsf{cost}(\pi_{\sigma_m}|\mu) > c$, so $\inf_{\sigma_a \in \Sigma_a} \mathbb{P}(\sigma_m, \sigma_a) \leq \mathbb{P}(\sigma_m, \sigma_\mu) < -c$. Since $c \in \mathbb{N}_0$ was arbitrary, thus $\inf_{\sigma_a \in \Sigma_a} \mathbb{P}(\sigma_m, \sigma_a) = -\infty$. Since $\sigma_m \in \Sigma_m$ was arbitrary, we are done.

Finally, we claim that $\inf_{\sigma_a \in \Sigma_a} \sup_{\sigma_m \in \Sigma_m} \mathbb{P}(\sigma_m, \sigma_a) = -\infty$. Indeed, let $c^* \in \mathbb{N}_0$ denote an arbitrary cost; we exhibit an adversary strategy $\sigma_{c^*}$ that guarantees a higher cost. First, define $\ell_0 \equiv 2^{3^{c^*}}$, and for each $c \in \{0, 1, ..., c^*-1\}$ define $\ell_{c+1} \equiv \lfloor \sqrt{\ell_c /2} \rfloor$. It is straightforward to verify by induction that for each $c \in \{0, 1, ..., c^*\}$, we have $\ell_c \geq 2^{2^{c^*-c}}$, and in particular $\ell_{c^*} \geq 2$. We construct $\sigma_{c^*}$ sequentially: at each adversary history $h$ such that we have already defined the adversary's actions at all earlier adversary histories, define $\sigma_{c^*}(h)$ as follows.
\begin{itemize}
\item If (i)~$\mathbb{D}(h) \neq \emptyset$, (ii)~there have thus far been $c \in \{0, 1, ..., c^*-1\}$ queries, and (iii)~there is an $\ell_c$-deficient partition record $(\mathcal{P}, \alpha)$ that extends the record given by all previous responses, then if $\mathsf{q}$ denotes the most recent (and only unanswered) query, \hyperlink{Lemma4}{Lemma~4} promises there is $(\mathcal{P}^+, \alpha^+) \in \mathcal{R}(\mathsf{q}|\mathcal{P}, \alpha)$ that is $\ell_{c+1}$-deficient. Select a measure $\mu \in \mathbb{D}$ associated with this ultraresponse, then select the response $\mathsf{r}(\mathsf{q}|\mu)$.

\item Otherwise, select any response.
\end{itemize}
This completes our definition of $\sigma_{c^*}$. By construction, for each mediator strategy $\sigma_m \in \Sigma_m$ and each $c \in \{0, 1, ..., c^*\}$, if there is a mediator history $h$ at which $c$ queries have been already asked that is on-path given $(\sigma_m, \sigma_{c^*})$, then the record of all previous responses can be extended to a partition record $(\mathcal{P}, \alpha)$ that is $\ell_c$-deficient, and moreover $\ell_c \geq \ell_{c^*} \geq 2 > 1$, so there is no allocation that is $e$-proportional for each member of $\mathbb{D}(\mathcal{P}, \alpha)$ and $\mathbb{D}(\mathcal{P}, \alpha) \subseteq \mathbb{D}(h)$. Since $c \in \{0, 1, ..., c^*\}$ was arbitrary, thus $\mathbb{P}(\sigma_m, \sigma_{c^*}) < -c^*$. Since $\sigma_m \in \Sigma_m$ was arbitrary, thus $\inf_{\sigma_a \in \Sigma_a} \sup_{\sigma_m \in \Sigma_m} \mathbb{P}(\sigma_m, \sigma_a) \leq \sup_{\sigma_m \in \Sigma_m} \mathbb{P}(\sigma_m, \sigma_{c^*}) < -c^*$. Since $c^* \in \mathbb{N}_0$ was arbitrary, we are done.~$\blacksquare$

\phantomsection

\end{document}